# An Automatic Genetic Algorithm Framework for the Optimization of Three-dimensional Surgical Plans of Forearm Corrective Osteotomies

Fabio Carrillo* [a,b], Simon Roner [a,c], Marco von Atzigen [a,b], Andreas Schweizer [a,c], Ladislav Nagy [a,c], Lazaros Vlachopoulos [a,c], Jess G. Snedeker [b], Philipp Fürnstahl [a]

[a] Computer Assisted Research and Development Group, Balgrist University Hospital, University of Zurich, Forchstrasse 340, CH-8008 Zurich, Switzerland

[b] Laboratory for Orthopaedic Biomechanics, Institute for Biomechanics, ETH Zürich, Balgrist Campus, Lengghalde 5, CH-8008 Zurich, Switzerland

[c] Department of Orthopaedics, Balgrist University Hospital, University of Zurich, Forchstrasse 340, CH-8008 Zurich, Switzerland

**\* Corresponding Author**:

**Fabio Carrillo, Msc ETH**

Computer Assisted Research and Development Group, Balgrist University Hospital

University of Zurich, Forchstrasse 340, CH-8008 Zurich, Switzerland

E-mail: fabio.carrillo@balgrist.ch

E-mail addresses:

fabio.carrillo@balgrist.ch; simon.roner@balgrist.ch; marco.vonatzigen@balgrist.ch; andreas.schweizer@balgrist.ch; ladislav.nagy@balgrist.ch; lazaros.vlachopoulos@balgrist.ch; jess.snedeker@balgrist.ch; philipp.fuernstahl@balgrist.ch

## Abstract

Three-dimensional (3D) computer-assisted corrective osteotomy has become the state-of-the-art for surgical treatment of complex bone deformities. Despite available technologies, the automatic generation of clinically acceptable, ready-to-use preoperative planning solutions is currently not possible for such pathologies. Multiple contradicting and mutually dependent objectives have to be considered, as well as clinical and technical constraints, which generally require iterative manual adjustments. This leads to unnecessary surgeon efforts and unbearable clinical costs, hindering also the quality of patient treatment due to the reduced number of solutions that can be investigated in a clinically acceptable timeframe. In this paper, we propose an optimization framework for the generation of ready-to-use preoperative planning solutions in a fully automatic fashion. An automatic diagnostic assessment using patient-specific 3D models is performed for 3D malunion quantification and definition of the optimization parameters' range. Afterward, clinical objectives are translated into the optimization module, and controlled through tailored fitness functions based on a weighted and multi-staged optimization approach. The optimization is based on a genetic algorithm capable of solving multi-objective optimization problems with non-linear constraints. The framework outputs a complete preoperative planning solution including position and orientation of the osteotomy plane, transformation to achieve the bone reduction, and position and orientation of the fixation plate and screws. A qualitative validation was performed on 36 consecutive cases of radius osteotomy where solutions generated by the optimization algorithm (OA) were compared against the gold standard solutions generated by experienced surgeons (Gold Standard; GS). Solutions were blinded and presented to 6 readers (4 surgeons, 2 planning engineers), who voted OA solutions to be better in 55% of the time. The quantitative evaluation was based on different error measurements, showing average improvements with respect to the GS from 20% for the reduction alignment and up to 106% for the position of the fixation screws. Notably, our algorithm was able to generate feasible clinical solutions which were not possible to obtain with the current state-of-the-art method.

## Research Highlights

- Automatic diagnosis strategy based on bony landmarks.
- Automatic placement of the fixation plate.
- Two-stage weighted multi-objective optimization based on a genetic algorithm
- Novel bone protrusion evaluation considering bone contact and surfaces gaps.
- Patient-specific screw optimization based on bone density information.
- Capability of considering all types of common osteotomies: single-cut, opening wedge, closing wedge.

# Abbreviations

2D: Two-dimensional

3D: Three-Dimensional

CA: Computer-assisted

CAD: Computer-aided design

CASPA: Planning software developed at the Computer Assisted Research and Development Group

CT: Computed tomography

DoF: Degrees of freedom

GPU: Graphics Processing Unit

GS: Gold standard

OA: Optimization algorithm

SSM: Statistical shape models

ROM: Range of motion

# 1. Introduction

Post-traumatic healing in non-anatomical positions (malunions) or congenital deformations of bones can cause limitations in the range of motion (ROM) of the patient, generate pain and, if not treated properly, result in severe degenerative pathologies such as osteoarthritis (Nagy et al., 2008). The current gold standard for surgical treatment of these pathologies is the restoration of the normal anatomy in a surgical procedure known as a corrective osteotomy. In this procedure, the pathological bone is cut into two or more fragments, the fragments are realigned (clinical term: reduced) to their physiological position and stabilized with an osteosynthesis implant (Schweizer et al., 2010). However, the correction of bone malunions is highly patient-specific, requiring performing a complex 6-degree-of-freedom (DoF) correction (rotation and translation) for each bone fragment, in order to restore the physiological anatomy of the patient. Therefore, corrective osteotomies are technically challenging to perform without careful diagnosis and detailed preoperative planning of the procedure. Moreover, the successful outcome of a corrective osteotomy depends also on precise intraoperative navigation of the bone reduction (Fürnstahl et al., 2016).

Conventional two-dimensional (2D) preoperative planning approaches based on X-ray and Computed Tomography (CT) images fail to correctly assess the inherently three-dimensional (3D) nature of bone malunions (Schweizer et al., 2010). Another drawback of 2D preoperative planning is that it cannot be used for surgical navigation, and surgeons must rely mainly on outdated, rudimentary surgical techniques to achieve the desired bone correction. Computer-assisted 3D preoperative planning addresses those problems. Several works have established 3D preoperative planning as the state-of-the-art technique for corrective osteotomies (Athwal et al., 2003; Dobbe et al., 2011; Fürnstahl, 2010; Fürnstahl et al., 2016), due to its clear benefits in patient treatment. It allows the quantification of malunions and its corrections in all 6 DoF. 3D preoperative planning offers also the possibility of precise pre-calculation of the osteotomy fixation using 3D representations of the fixation plates and fixation screws (Dobbe et al., 2011; Miyake et al., 2012a; Schweizer et al., 2010; Schweizer et al., 2013). Lastly, it enables the translation of the preoperative planning intraoperatively by means of surgical navigation, based either on patient-specific instruments (PSI) (Fürnstahl et al., 2016; Miyake et al., 2011; Murase et al., 2008; Omori et al., 2014) or on optical navigation systems (Andress et al., 2018). Thus, the introduction of 3D preoperative planning techniques marked a paradigm shift in patient treatment, allowing performing complex surgical procedures that would not be possible using conventional techniques (Athwal et al., 2003; Dobbe et al., 2014; Fürnstahl et al., 2016; Kunz et al., 2013; Roner et al., 2017; Schweizer et al., 2013; Schweizer et al., 2016; Zdravkovic and Bilic, 1990).

In our institution, the current state-of-the-art preoperative planning of long-bone osteotomies encompass the following steps: in a first step, patient-specific 3D triangular surface models (hereinafter: 3D models) of the bones are generated based on the segmentation of the CT data of the patient (Fürnstahl et al., 2008), which is part of our standard clinical procedure (Fürnstahl et al., 2008). After obtaining the patient-specific 3D bone models, a diagnosis of the malunion is performed by comparison of the pathological bone model to a reconstruction target (**Figure 1A**). Usually, a mirrored model of the contralateral healthy side is used as the reconstruction target. The deformity analysis allows the definition of the needed osteotomy cuts along the pathological bone model, which are simulated as shown in **Figure 1B**. The resulting bone fragments can then be realigned into their anatomical position by realigning the bone fragments to fit the reconstruction target (Fürnstahl, 2010; Fürnstahl et al., 2016; Murase et al., 2008). The final step of the preoperative planning procedure is the simulation of the fixation of the osteotomy by integrating 3D models of the fixation plate and screws (**Figure 1C**). In total, the planning process for corrective osteotomies requires the definition of 18 DoF, without including the orientation and position of the screws. This preoperative plan can be translated into the operation room by means of patient-specific navigation instruments, designed according to the preoperative plan and later 3D-printed, allowing the surgeons to perform intraoperatively step-by-step the previously simulated procedure (Fürnstahl et al., 2016; Omori et al., 2014; Rosseels et al., 2018).

Despite clear advantages of 3D preoperative planning, the generation of preoperative planning solutions requires the manual calculation of the aforementioned steps by trial and error, even when using dedicated 3D planning software (Fürnstahl et al., 2016; Roner et al., 2017; Schweizer et al., 2010). The development of a clinically feasible solution also requires close collaboration between surgeons, providing the clinical knowledge, and engineers, who have the technical expertise. As the availability of surgeons needed for consultancy is often very limited, the generation of an optimal preoperative solution for extra-articular long-bone osteotomies can add up to 4 hours (Fürnstahl et al., 2016) and might involve several iterations of manual adjustments. This incurs unwanted clinical costs as a consequence of the human workload spent on manual processes. Moreover, only a reduced number of clinically feasible solutions can be investigated due to the constrained clinical timeframe.

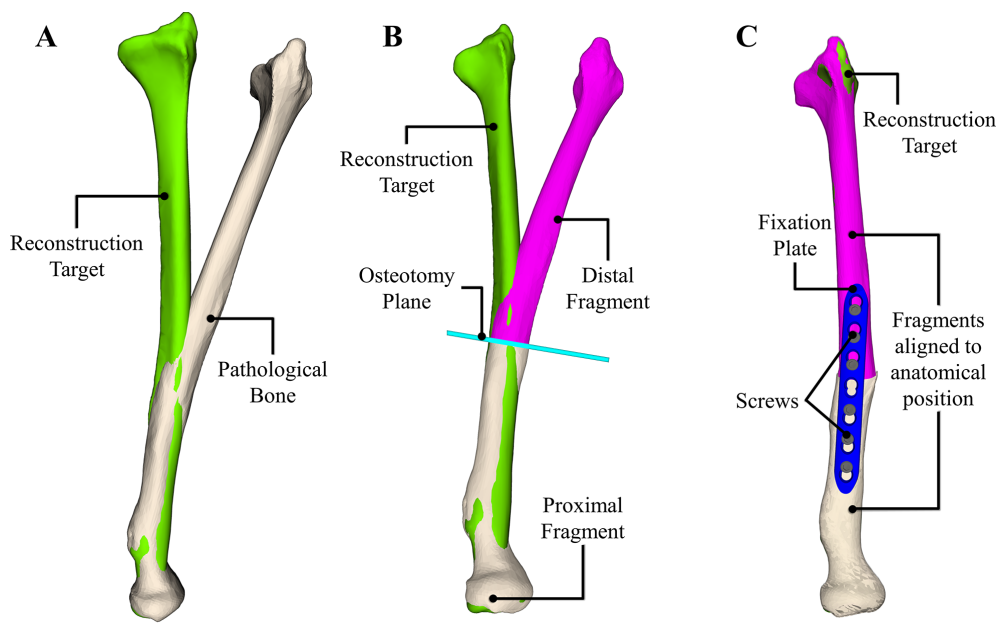

**Figure 1: Overview of state-of-the-art preoperative planning of long-bone osteotomies.** (A) 3D bone models obtained through segmentation of CT data. The pathological bone model is shown in white and the reconstruction target in green. Diagnosis of the malunion is done by comparison of the two bones. (B) An osteotomy plane (cyan) is created and the pathological bone model is cut, generating the proximal (white) and distal (magenta) bone fragments. (C) Bone reduction is simulated by aligning the generated fragments to the reconstruction target, and the osteotomy is fixated with a fixation plate (in blue) and its corresponding screws (in gray).

One possibility to overcome those challenges is the implementation of a computer-based planning approach, able to systematically and automatically generate optimal preoperative planning solutions. The automatic generation can significantly decrease clinical costs and reduce unnecessary interaction times of the collaborators. It could also improve the quality of patient treatment by considering a larger range of parameters and solutions in order to generate better preoperative planning solutions than those obtained by a human planner. Moreover, with the current trend of the health industry towards digitalization of patient data and treatment solutions, automatic methods become an essential asset for optimal clinical treatments.

However, the implementation of an automatic optimization approach is a challenging task. Early on, Zdravkovic and Bilic (1990) proposed a computer-assisted preoperative planning framework for orthopedic surgeries in order to handle the complex tasks involved in the preoperative planning process. The latter, includes multiple nonlinear, discontinuous and non-differentiable planning objectives, making their mathematical manipulation difficult. Some objectives are contradicting and tightly associated with each other, which can cause the worsening of one objective while trying to improve another one.

In the field of corrective osteotomies of long bones, only a few works have tried to tackle the automatic optimization problem (Belei et al., 2007; Carrillo et al., 2017; Schkommodau et al., 2005). Although these approaches have been promising and are pioneers in the filed of automatic optimization for surgical planning of orthopedic surgeries, they still lack of solutions that can be readily applied in clinical practice without further modifications.

In this paper, we present a multi-staged, multi-objective optimization approach based on the artificial intelligent methods for the generation of ready-to-use preoperative planning solutions. The system is capable of calculating solutions considering all common bone malunions, osteotomy types, and surgical approaches. We have introduce the following key contributions with respect to our previous work (Carrillo et al., 2017):

- An automatic diagnosis of bone deformation based on bony landmarks, which allows automatic narrowing of the optimization search space. In our previous work, the diagnosis of the bone deformation was assumed as given.

- Automatic placement of the fixation plate using information provided by statistical shape models (SSM), in contrast to the manual definition of the feasible fixation areas used in (Carrillo et al., 2017).

- A novel bone protrusion evaluation that supports the generation of better solutions by considering bone contact and surfaces gaps. The use of bone protrusion represents a more realistic clinical metric than the minimization of the cut surface used in our previous work.

- In contrast to the heuristic approach employed in our previous work, in this paper, we have developed a patient-specific screw purchase optimization based on bone density information extracted from CT data.

- The capability of considering all types of common osteotomies (single-cut, opening wedge, closing wedge) along the entire Radius, in order to generate solutions that would be difficult to achieve for a human planner within a reasonable time. The algorithm presented in (Carrillo et al., 2017) was only able to handle distal radius osteotomies and was not capable of generating single-cut solutions.

- We have also reduced the computational effort of the strategy presented in (Carrillo et al., 2017) by applying a different multi-stage approach.

- Finally, we have provided a profound clinical and mathematical explanation of all the optimization objectives, which was previously missing.

The developed optimization framework was validated clinically against state-of-the-art preoperative planning solutions on a consecutive series of patients with forearm malunions, previously treated at our institution. In the following, we will give a brief overview of existing preoperative planning approaches for orthopedic surgeries. In section 2, we describe in detailed the proposed optimization framework. In section 3, dataset, experimental set-up, and results are presented. In section 4 a discussion about results, impact, and limitations of this work is given. Finally, in section 5, we draw final conclusions and give an outlook about future work.

## 1.1. Related Work

The first step towards the generation of a complete 3D preoperative planning is the patient diagnosis done through the analysis of the bone malunion. In 3D preoperative planning, patient-specific 3D bone models generated from the segmentation of multi-planar data (Lorensen and Cline, 1987) are superimposed over a healthy reconstruction template using semi-automatic registration methods (Kawakami et al., 2002; Schenk et al., 2016; Schweizer et al., 2010; Vlachopoulos et al., 2017). Afterward, the bone deformation is quantified by means of clinical and mathematical measurements (Fürnstahl et al., 2016; Gosse et al., 1997; Murase et al., 2008; Subburaj et al., 2010), providing the basis for the preoperative planning of the surgical correction. Here, 4 main objectives have to be considered: osteotomy cut plane, reduction of the bone fragments, position of the fixation plate and direction and position of fixation screws.

Current state-of-the-art of 3D preoperative planning for corrective osteotomies uses dedicated planning software for the manual calculation of each of these objectives. We refer the interested readers to (Fürnstahl, 2010; Fürnstahl et al., 2016; Schweizer et al., 2016; Vlachopoulos et al., 2015) for more information about the 3D planning tool. The latter facilitates the process of generation of a 3D preoperative plan, however the basic primitives operations needed for creating a preoperative plan are cutting Boolean operations and the possibility for interactive transformation, which would be also available in any dedicated CAD software.

The process of creating each of the steps involved in the preoperative plan is difficult to control even by skilled engineers, as any change done to one of the parameters (e.g., position of fixation screws, inclination of osteotomy plane) must be manually propagated across the different objectives (Athwal et al., 2003; Belei et al., 2007; Bilic et al., 1994; Carrillo et al., 2017; Fürnstahl, 2010). Current-state-of-the-art planning (Fürnstahl et al., 2016; Roner et al., 2017; Vlachopoulos et al., 2015) is also incapable of handling contradicting objectives, meaning for example that an improvement in the accuracy of the bone reduction can subsequently deteriorate the position of the fixation plate or generate solutions with non-feasible osteotomy cuts. Similarly, an improvement in the position of the osteotomy cut can cause a less fitting position for the placement of the fixation plate.

Existing automatic methods have only solved a reduced problem set, failing to include all objectives needed for a ready-to-use clinical solution , i.e., osteotomy cut plane, reduction of the bone fragments, position of the fixation plate and direction and position of fixation screws. Such is the case of previously described methods (Eck et al., 1990; Menetrey and Paul, 2004), where an automatic calculation was performed only for the osteotomy plane. Eck et al. (1990) calculated the osteotomy plane in femoral head reduction planning using a nonlinear optimization algorithm, based on a least-square-approximation solver. Menetrey and Paul (2004) did a similar parametrization of the osteotomy plane and wedge size for osteotomies around the knee, optimizing only the reduction alignment.

Schkommodau et al. (2005) developed a multi-objective optimization strategy for corrective osteotomies of lower extremities that considered the following optimization objectives: leg length, translation, antetorsion, and angulation. Their method solved a simplified osteotomy sub-problem with only 12 DoF, which did not include the position of the fixation plate or the fixation screws into the optimization process. The multi-objective problem was solved by a sequential quadratic programming algorithm and considered the influence of all these objectives in the preoperative planning. The approach was later extended by Belei et al. (2007) to account for different osteotomy types (closing wedge, opening wedge, single cut) and to consider also double osteotomy solutions. These approaches represent the first attempt to automate the planning of corrective osteotomy. However, the planning was based on simplified geometry rather than on patient anatomy and it relied on intraoperatively calibrated fluoroscopic datasets.

Learning-based methods have proven to be effective in similar applications of medical image processing techniques (Criminisi and Shotton, 2013; Esfandiari et al., 2018; Tschannen et al., 2016). In the field of shoulder arthroplasty, Tschannen et al. (2016) presented an automatic algorithm for preoperative planning of the resection plane for arthroplasty of the proximal humeral head, based on random regression forests. The approach allowed controlling the orientation, position, and size of the prosthetic humeral head in relation to the humeral shaft, using a direct mapping between the CT image and the parameters of the resection plane. The estimation of the plane position using CT data could be of interest to speed-up the convergence of optimization algorithms. Another interesting application is found in the field of spine surgery, where the problem of pedicle screw placement and pose estimation has been extensively studied (Farshad et al., 2017; Scheufler et al., 2011). Also, Esfandiari et al. (2018) proposed an algorithm based on convolutional neural networks for the 6-DoF estimation of the screw position and direction using intraoperative fluoroscopy data and estimation of the bone density. Similar approaches, taking into account the bone density information of the patient, should be considered in long-bone osteotomies to increase the quality of the osteotomy fixation.

To the best of our knowledge, only our previous work (Carrillo et al., 2017) deals with an 18-DoF optimization problem for the generation of preoperative planning solutions of corrective osteotomies. The optimization approach presented in (Carrillo et al., 2017) has been taken as the basis for the core optimization algorithm (section 2.3.3) of the herein presented framework.

## 2. Methods

We have developed an optimization framework for the generation of ready-to-use preoperative planning solutions for corrective osteotomies. A complete overview of our approach is given in **Figure 2**. The framework receives the 3D bone models of the patient, the reconstruction target and the fixation plate as an input (section 2.1). Afterward, an automatic diagnosis of the malunion is performed by identification of the pathological area, encoding also feasible plate regions and associated clinical constraints (section 2.2). This information is used by the optimization module, where the generation and optimization of preoperative planning solutions take place (i.e., the simulation of the osteotomy cut, realignment of the fragment and virtual fixation of the osteotomy) using a multi-stage optimization strategy (section 2.3). A genetic algorithm approach is used for optimization in which the objectives are encoded in a real-valued chromosome. The optimization is controlled by means of fitness evaluation functions, which provide quality measurements for the attainment of the different objectives. The output of the framework is a complete preoperative planning solution, including position and orientation of the osteotomy plane, the 6-DoF transformation to achieve the reduction, and the position and orientation of the fixation plate and screws. Our approach has been designed to cover all long-bone malunions, however, we have focused the implementation of our method on the radius due to the high rate of radius osteotomy procedures performed yearly at our institution.

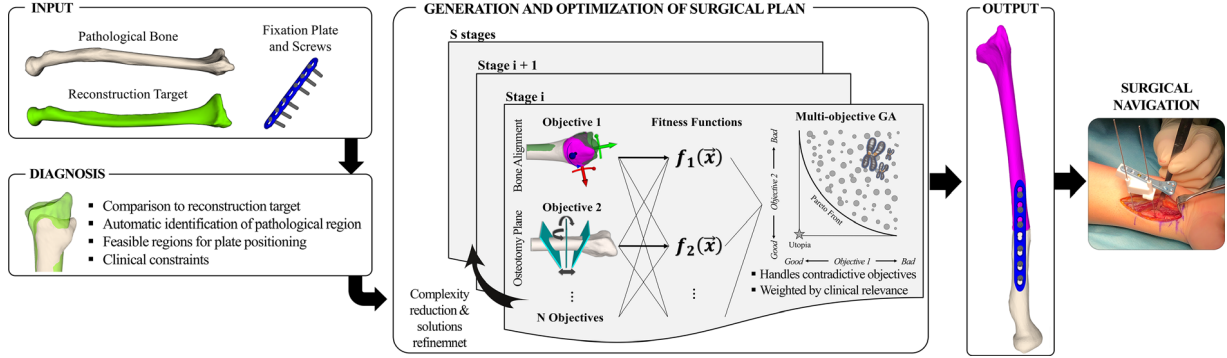

**Figure 2:** Overview of the automatic optimization framework for the 3D planning of long bone corrective osteotomy surgeries.

## 2.1. Input data generation

In our institution, patient-specific 3D models of bone anatomy are used in the standard treatment workflow for complex malunions. For a better understanding, we briefly describe the model generation process although it is not an integral part of our optimization framework. Patient-specific 3D bone models of the forearm were generated from CT data (slice thickness 1 mm; 120 kV; Philips Brilliance 64 CT, Philips Healthcare, The Netherlands) using thresholding and region-growing algorithms of commercial segmentation software (Mimics® Medical, Version 19.0, Materialise 2016, Leuven, Belgium). 3D models are generated using the Marching Cube algorithm (Lorensen and Cline, 1987) and given in the form of triangular surface meshes (stereolithographic models; STL) as described by (Roscoe, 1988). In our institution, the mirrored model of the contralateral bone is always used as the anatomical (healthy) reconstruction target. As the last input parameter, a feasible osteosynthesis implant (fixation plate) must be chosen, used for the fixation of the bone fragments after osteotomy.

All 3D models were transformed into a common anatomical reference frame. In the case of the radius, we have used the anatomical coordinate system described in (Vlachopoulos et al., 2015). The coordinate system is denoted by $\overrightarrow{CS}$ and it is oriented as shown in **Figure 3A**. Different anatomical and clinically relevant landmarks regions were defined and used throughout the entire pipeline. 7 point sets of landmarks regions, of 5 mm radius size, were automatically generated on the distal parts of both, pathological radius $\{LP_1, ..., LP_7\}$ and reconstruction target $\{LR_1, ..., LR_7\}$ as depicted on **Figure 3B**. In our implementation, each set $LP_l$ and $LR_l$ contains a total of $N = 50$ points, and each point set $LP_l$ is in correspondence to the reciprocal point set $LR_r$.

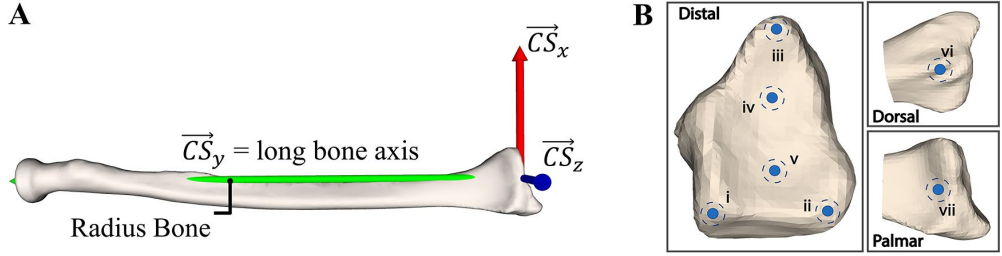

**Figure 3: Anatomical Axis and Bony Landmarks.** (A) Anatomical coordinate system; coordinate axis $\vec{CS}_x$ is depicted by the red arrow and represents the radioulnar direction, the coordinate axis $\vec{CS}_y$ is depicted by the green arrow and represents the axial direction (long bone axis), and coordinate axis $\vec{CS}_z$ is shown in blue and indicates the dorsal-palmar direction. (B) Anatomical landmarks of the distal radius: (i) dorsal distal edge of the sigmoid notch; (ii) palmar distal edge of the sigmoid notch; (iii) center of distal radius styloid; (iv) center of scaphoid facet; (v) center of lunate facet; (vi) Lister's tubercle and (vii) center of radius' watershed line.

## 2.2. Automatic Diagnosis of Bone Malunion

A common clinical practice for the 3D diagnosis of a bone malunion is to align the healthy proximal and distal joints of the pathological bone to the corresponding regions of the reconstruction target (Schweizer et al., 2010b). This process allows obtaining a visual approximation of the deformed area of the pathological bone and it defines also the most proximal and most distal regions of the bone at which an osteotomy cut can be performed. The bone malunion can then be quantified in 3D by the relative $4x4$ transformation matrix between the proximally and distally aligned bone models (**Figure 4**). The transformation matrix encodes the translation and rotational deviation of the malunion in all 6 DoF with respect to the anatomical coordinate system $\vec{CS}$ (Schweizer et al., 2010; Vlachopoulos et al., 2017). Based on this well-established principle, we have developed an automatic method for the calculation of the deformity region, the deformity profile, and the quantification of the bone malunion.

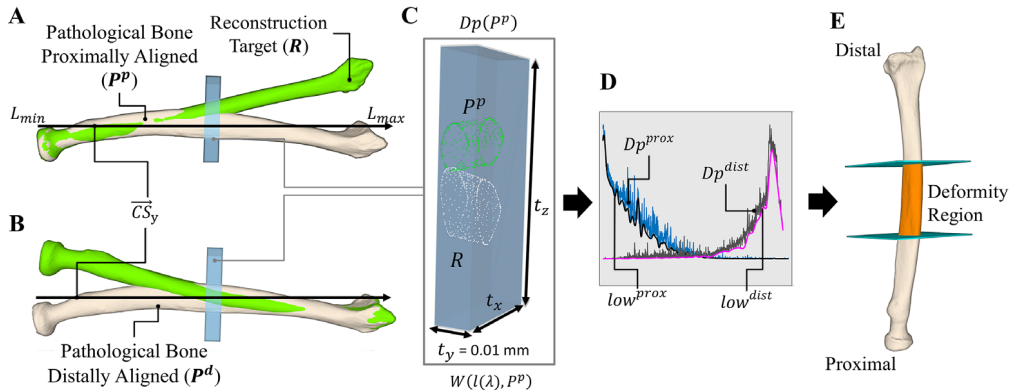

**Figure 4: Method for automatic diagnosis of malunions.** (A-B) The pathological bone **P** is proximally (A) and distally (B) aligned to the reconstruction target **R**. (C) Example calculation of windows function for $P^p$; function $Dp$ is calculated along the $\vec{CS}_y$ of the bone, from $L_{min}$ to $L_{max}$, with window function $W(\cdot)$ of thickness $t = (1\ mm, \infty, \infty)$. (D) Deformity curves; Lower envelopes $low^{prox}$ (bold black signal) of $Dp(P^{prox})$ (blue signal), and $low^{dist}$ (bold magenta signal) of $Dp(P^{dist})$ (gray signal), are calculated, corresponding to both proximal and distal alignments. (E) Deformity region is shown in orange, between the most proximal and most distal planes obtained from the deformity threshold $d_{Th}$ applied to $low^{prox}$ and $low^{dist}$.

In the automatic diagnosis, the pathological bone model **P** is coarsely aligned to the reconstruction target **R** using a landmark-based registration method, where each $LP_l$ of the pathological bone is registered to its reciprocal $LR_l$ of the reconstruction target, using standard iterative closest point (ICP) (Besl and McKay, 1992). Afterwards, the minimum and maximum points $L_{min}$ and $L_{max}$ are respectively calculated, on the bone length axis $\vec{CS_y}$ with respect to $P$. Let

$$\delta(x, Y) = \arg\min_{y_j \in Y} \|x - y_j\| \quad \text{(Eq. 1)}$$

be the nearest neighbor function yielding the point $y_j$ of a point set or line $Y$ with the smallest Euclidean distance to a point $x$. $L_{min}$ and $L_{max}$, and the line segment $L(\lambda)$ between them can then be defined as

$$L_{min} = \delta(p_{min}, \vec{CS_y}) \mid p_{min} = \min_{p_i \in P}(p_i \cdot \vec{CS_y}), \qquad L_{max} = \delta(p_{max}, \vec{CS_y}) \mid p_{max} = \max_{p_i \in P}(p_i \cdot \vec{CS_y}),$$

$$\text{and } L(\lambda) = L_{min} + \lambda(L_{max} - L_{min}) \mid 0 \leq \lambda \leq 1,$$

where $\cdot$ denotes the scalar product.

Following the recommendation of (Vlachopoulos et al., 2017) for the size of the proximal and distal alignment regions, $L(\lambda)$ is used to determine the points of $P$ that lie in the most proximal and distal 30% with respect to $\vec{CS_y}$:

$$\{ p_i \mid p_i \cdot \vec{CS_y} \leq L(0.3) \cdot \vec{CS_y} \} \text{ and } \{ p_i \mid \|p_i \cdot \vec{CS_y} - L_{min}\| \leq \|L(0.3) - L_{min}\|\}$$

These points are then used to align the proximal and distal joints of **P** to **R** (**Figure 4A**). The alignment is done using ICP registration (Besl and McKay, 1992) and the proximally and distally aligned models are denoted by $\boldsymbol{P^p}$ and $\boldsymbol{P^d}$, respectively.

Afterward, the deformity profiles $Dp(P^p)$ (**Figure 4D**, blue signal) and $Dp(P^{dp})$ (**Figure 4D**, grey signal) are obtained from the deformity function

$$Dp(PS) = \sqrt{\frac{1}{|PS|} \sum_{\substack{i=1,\dots,|PS| \\ \lambda=0,0.1,\dots,1}} W(l(\lambda), p_i)^2 - \delta(W(l(\lambda), p_i), W(l(\lambda), R))^2}, \quad \text{(Eq. 2)}$$

where $PS$ is a pointset, $|PS|$ is the cardinality of the pointset and $W(\cdot)$ denotes the window function

$$W(c, PS) = \begin{cases} \{ p_i \in PS \mid c - t < p_i < c + t \} \\ \emptyset : \quad otherwise \end{cases}. \quad \text{(Eq. 3)}$$

$c$ is the center of the window and $t$ is an adjustable user-defined thickness. As shown in **Figure 4C,** we have used a thickness $t_y = 0.01 \, mm$, and we have taken $t_x$ and $t_z$ to be large enough to enclose **P** and **R** along $\vec{CS_x}$ and $\vec{CS_z}$.

In order to reduce noise in the signals, we have implemented filtering using the lower envelopes $low^{prox}$ (**Figure 4D**, bold black line) and $low^{dist}$ (**Figure 4D**, bold magenta line) of $Dp(P^p)$ and $(P^d)$, respectively. Based on $low^{prox}$ and $low^{dist}$, a deformity threshold $d_{Th} = 10\%$ defines the part of the bone in which the deformity profile deviates more than 10% from their minimum value (**Figure 4E**). This region defines the most proximal and most distal sites for feasible osteotomy cuts.

Finally, the malunion is quantified by a 4x4 homogenous transformation matrix $M_{mal}$ representing the difference between the proximally and distally aligned **P**. $M_{mal}$ contains the information about the rotational and translational error of the bone malunion with respect to $\overrightarrow{CS}$. The rotational deviation along each direction of $\overrightarrow{CS}$ is obtained by applying the Euler-Rodrigues formula (Schneider and Eberly, 2002) to the rotational component of $M_{mal}$, which yields the Euler angles $\varphi_x, \varphi_y$ and $\varphi_z$. The translational deviation along each coordinate axis is obtained from the translational parts $T_x, T_y$ and $T_z$ of $M_{mal}$ (Schneider and Eberly, 2002). Thus, the 6-DoF quantification of bone malunion is given by the vector $\vec{V}_{mal} = (\varphi_x, \varphi_y, \varphi_z, T_x, T_y, T_z)$.

## 2.3. Multi-stage optimization

Our optimization approach comprises three parts. First, each clinical goal is characterized mathematically by means of optimization parameters, boundaries, and constraints. These parameters are used for the automatic generation of solutions at each optimization step, and the boundaries and constraints are used to control the search space of the algorithm. Our mathematical translation of the clinical goals into optimization objectives is explained in section **2.3.1**. Secondly, the optimization algorithm requires a way to evaluate the performance of the optimization objectives. In our algorithm, this evaluation is performed by tailored fitness functions, developed for the minimization or maximization of the optimization parameters. The appropriate definition of the fitness function is crucial for the generation of feasible solutions and convergence of the algorithm. We give a detailed description of the fitness functions in section **2.3.2**. Lastly, an optimization strategy must be chosen for the generation of solutions. In our case, the optimization algorithm had to be able to handle multiple objectives with 18 DoF and nonlinear constraints. We have implemented a multi-stage optimization approach based on the multi-objective genetic algorithm NSGA-II (Deb et al., 2000), further explained in section **2.3.3**.

### 2.3.1. Optimization Objectives

A complete preoperative solution for long bone osteotomies entails the calculation of 4 clinical objectives, as shown in **Figure 5**: (A) The position and orientation of the osteotomy cut plane (6 DoF), (B) the translation and rotation of the generated bone fragments for the reduction to the reconstruction target (6 DoF), (C) the position and orientation of the fixation plate (6 DoF) and (D) the position and orientation of the plate's screws into the reduced fragments. The definition of each of these objectives requires more than a simple mathematical translation into adjustable parameters. We have to carefully decide on the parameters to control such that the parameter space can be minimal, and the controllability of the objectives can be directly evidence from a parameter change. In this section, we describe the strategy on the mathematical parameters, constraints, and boundaries associated with each objective.

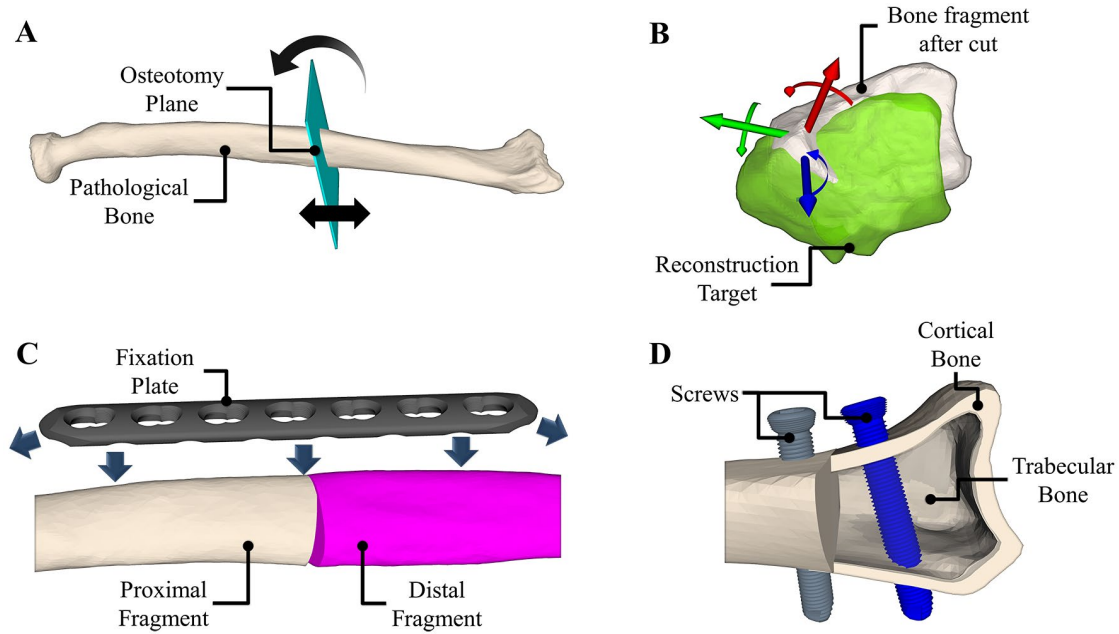

**Figure 5: Optimization objectives.** (A) Position and orientation of the osteotomy cut plane (in cyan) along the pathological bone (in white). (B) Reduction alignment of the generated bone fragments (in white) with respect to the reconstruction target (in green). (C) Position of the fixation plate (in gray) with respect to the proximal (in white) and transformed distal fragment (in magenta). (D) Position and orientation of the screws of the fixation plate (in blue and gray), used for the stabilization and fixation of the bone fragments.

A. *Osteotomy Plane* (**Figure 5A**). The osteotomy plane determines the size and location of the osteotomy and the orientation of the bone cut. It is represented by the position $pl$ (3 DoF) and normal $\vec{n}_{pl}$ (3 DoF), with a total of 6 DoF. Regarding $pl$, we optimize only the position $pl_y$ relative to the long bone axis $\vec{CS}_y$ (**Table 1**). In extra-articular osteotomies, only the translation of the osteotomy plane along the bone length axis $\vec{CS}_y$ has an effect on the osteotomy, thus the translation of the osteotomy plane with respect to $\vec{CS}_x$ and $\vec{CS}_z$ is redundant. Together with the reduction alignment, the osteotomy plane determines the type of osteotomy and the shape of the wedge between the generated fragments. In our approach, we have considered all common osteotomy types (**Figure 6**): opening-wedge, closing wedge and single-cut osteotomy. **Figure 6A** shows the situation of the bone previous to realignment and after performing the osteotomy cut (in cyan).

In an opening-wedge osteotomy (**Figure 6B**) a bone cut is made and the generated bone fragments are reduced to their anatomical position, yielding a wedge-shaped opening. A too large gap may prevent healing or may result in implant failure due to instability. To avoid this, large gaps are filled with structural bone graft in the surgery to support healing (Fernandez, 1982; Fürnstahl et al., 2016; Murase et al., 2008). Oppositely, a closing-wedge osteotomy (**Figure 6C**) refers to an osteotomy where a bone wedge has to be removed to complete the bone reduction. This may occur due to an overlap between the two generated bone fragments after their reduction to the reconstruction target (Fürnstahl et al., 2016; Murase et al., 2008). Our algorithm calculates first the necessary reduction of the bone fragments and afterward the overlap is defined by simulating a second planar cut. Additionally, our approach has also the feature of calculating single-cut osteotomies (**Figure 6D**) (Dobbe et al., 2017; Sangeorzan et al., 1989). A single cut osteotomy is always preferred against closing wedge solutions due to better bone contact and faster bone healing (Merle d'Aubigné and Descamps, 1952; Roner et al., 2017; Sangeorzan et al., 1989). This type of osteotomy is technically more complex to plan manually, as the bone reduction is achieved by sliding and rotating the generated bone fragments along the osteotomy cut plane (Sangeorzan et al., 1989). We achieve the support of single-cut osteotomies in our framework by constraining the closing wedge osteotomy planes to be parallel and overlying. In this case, the wedge is limited to a maximum overlap of 1 mm along the $\overrightarrow{CS_y}$.

In each optimization stage, we use the polygon clipping algorithm of (Vatti, 1992) to cut model **P** by plane $(pl, n_{pl})$, yielding potential candidates of the proximal and distal fragments $\textbf{\textit{P}}^{prox}$ and $\textbf{\textit{P}}^{dist}$, respectively. The type of osteotomy is controlled by the maximum and minimum allowed wedge size using nonlinear constraints.

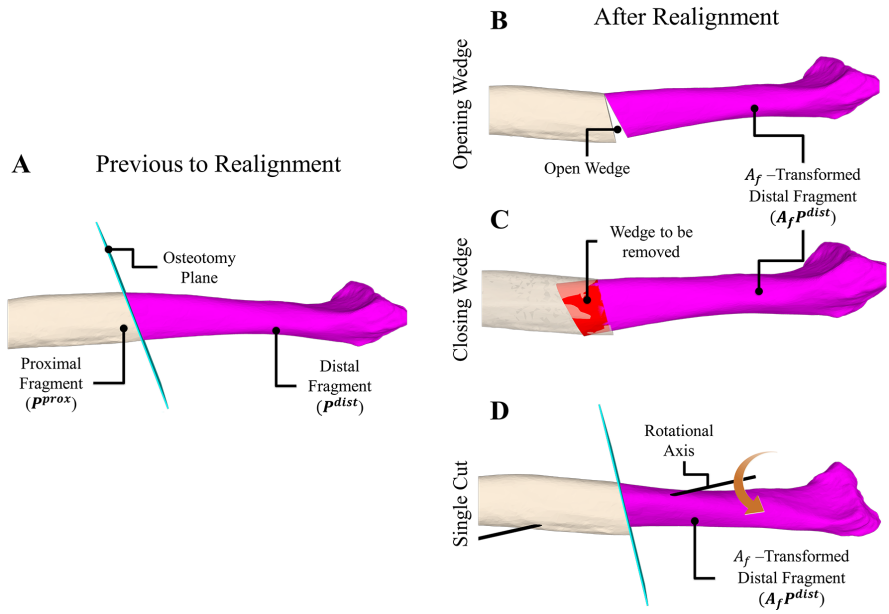

**Figure 6: Osteotomy Types.** (A) Situation after performing the osteotomy cut (shown in cyan) and before reduction to the reconstruction target. The proximal fragment is shown in white and the distal fragment in magenta. (B) Opening wedge: distal fragment (magenta) has been reduced to the reconstruction target, generating an opening between both fragments. (C) Closing wedge: in order to complete the osteotomy, after simulated reduction of the distal fragment (magenta), a second planar cut must be done to remove the wedge of the overlapping bone region (shown in red). (D) Single cut: reduction of the distal fragment is achieved by fragment rotation and translation along the osteotomy plane.

B. *Reduction Alignment* (**Figure 5B**). Once the osteotomy plane cut is executed, $\textbf{\textit{P}}^{dist}$ must be rotated and translated to match the given reconstruction target. This process of bone reduction represents one of the most important objectives for clinicians, because the precision of the reduction significantly influences the surgical outcome. Mathematically, the reduction alignment can be represented by a 4x4 homogenous transformation matrix ($A_f$), constructed from the rotation $R_f = (\varphi_x, \varphi_y, \varphi_z)$ and translation $\vec{T}_f = (T_{fx}, T_{fy}, T_{fz})$ parameters along each axis (6 DoF; **Table 1**).

C. *Position of Fixation Plate* (**Figure 5C**). Another factor influencing the success of the treatment is the position of the fixation plate. A poorly fitting plate results in a less controlled procedure, decreases stability of the reduction and even leads to inaccurate reduction, delayed bone healing or failure of the fixation plate (Blecha et al., 2005; Fürnstahl et al., 2016). Mathematically, the position of the fixation plate along the bone entails a 6 DoF task, which can be encoded in the optimization by a 4x4 homogenous transformation matrix

$A_p$. The latter is constructed from the rotation $R_p = (\theta_x, \theta_y, \theta_z)$ and the translation $\vec{T}_p = (T_{px}, T_{py}, T_{pz})$ of the fixation plate (**Table 1**), relative to the reference coordinate system $\overrightarrow{CS}$.

D. ***Screw Position and Orientation* (Figure 5D).** The quality of the solution is also considerably dependent on the position and orientation of the screws, which are required to achieve a stable bone-plate interface. Several studies have shown how the length of the fixation screw can affect the stability of the osteotomy and how poorly placed screws could lead to soft-tissue inflammation and ligament tearing (Blecha et al., 2005; Dougherty et al., 2008; Reitman et al., 2004). Mathematically, the position and orientation of a fixation screw $S_i$ is directly related to the position and orientation of the fixation plate, and can be described by

$$S_i = \vec{h}_{s_i} A_p (R_{s_i} + \vec{t}_{s_i}) \quad \text{(Eq. 4)}$$

where $\vec{h}_{s_i}$ is the screw axis, $A_p$ is the $4x4$ homogenous transformation matrix of the fixation plate position, $R_{s_i}$ is a $4x4$ rotation matrix encoding orientation of the screw relative to the plate and $\vec{t}_{s_i}$ is a constant translation vector containing the relative position of the screw hole with respect to the plate (**Table 1**). For uniaxial locking screw plate systems, where the orientation of the screw is already given by the manufacturer, $R_{s_i}$ is assumed to be constant, i.e., $R_{s_i} = \overrightarrow{dir}_{s_i} I_{4x4}$, where $\overrightarrow{dir}_{s_i}$ is a constant direction vector. In the optimization, the orientation of each screw is described by a vector $\vec{\theta}_{s_i}$, whose components are the Euler angles derived from $R_{s_i}$. The number of vectors $\vec{\theta}_{s_i}$ is determined by the number of screws $N_{si} = |\{s_i\}|$ of the fixation plate.

Once the optimization objectives have been defined, the optimization parameters are encoded directly as decimal values into a parametrization vector. The vector is further denoted as chromosome $\vec{x}$, defined as

$$\vec{x} = \begin{bmatrix} pl_y & n_{pl_x} & n_{pl_y} & n_{pl_z} & \varphi_x & \varphi_y & \varphi_z & T_{fx} & T_{fy} & T_{fz} & \theta_x & \theta_y & \theta_z & T_{px} & T_{py} & T_{pz} & \vec{\theta}_{s_1} & \vec{\theta}_{s_2} & \dots & \vec{\theta}_{s_{N_{si}}} \end{bmatrix}, \quad \text{(Eq. 5)}$$

which will be used by the fitness functions to control changes in the optimization objectives.

**Table 1:** Summary of optimization objectives and their associated fitness functions, optimization parameters and constraints

| Optimization Objective | Challenges | Fitness Function | Optimization Parameters | Parameters | Constraints |
|---|---|---|---|---|---|
| **Reduction Alignment** | ▪ Accuracy of joint surface<br>▪ Dependent on landmark weighting | Landmark-based Registration Error ($f_1$) | **Represented by $A_f$;**<br>▪ Rotation $R_f$ ($\varphi_x$, $\varphi_y$, $\varphi_z$)<br>▪ Translation $\vec{T}_f$ ($T_{fx}$, $T_{fy}$, $T_{fz}$) | 6 | Maximal allowed reduction error |
| **Osteotomy Plane** | ▪ Avoid longitudinal and intraarticular cuts<br>▪ Osteotomy type and location malunion influences reduction alignment | Bone Protrusion ($f_2$) | ▪ Position $pl$ ($pl_x, pl_y, pl_z$)<br>▪ Normal $\vec{n}_{pl}$ ($n_{pl_x}, n_{pl_y}, n_{pl_z}$) | 6 | ▪ For single-cut $\rightarrow \vec{n}_{pl}^{Pre} \parallel \vec{n}_{pl}^{Post}$<br>▪ Size of bone wedge<br>▪ Range given by diagnosis of section 2.2 |
| **Position Fixation Plate** | ▪ Gaps / steps between bone fragments and plate to be avoid<br>▪ Stable alignment<br>▪ Various plate types | Distance Fixation Plate - Bone Fragments ($f_3$) | **Represented by $A_p$;**<br>▪ Rotation $R_p$ ($\theta_x$, $\theta_y$, $\theta_z$)<br>▪ Translation $\vec{T}_p$ ($T_{px}, T_{py}, T_{pz}$) | 6 | ▪ Bone plate penetration<br>▪ Maximal allowed distance to bone<br>▪ Osteotomy site: palmar, dorsal, lateral. |
| **Position and Orientation of Screws** | ▪ Proximity to joint area<br>▪ Avoid penetration with osteotomy plane<br>▪ Patient-specific bone density | Screw Purchase ($f_4$) | $[\vec{\theta}_{s_1} \ \vec{\theta}_{s_2} \ \ldots \ \vec{\theta}_{s_{N_{si}}}]$<br>$\vec{\theta}_{s_i} = (\theta_{s_{i_x}}, \theta_{s_{i_y}}, \theta_{s_{i_z}})$<br>for $i = 1, \ldots, N_{si}$<br>w.r.t plate coordinate system and $A_p$ | 3 x $N_{si}$ ($N_{si}$ is the number of fixation screws) | ▪ Minimal distance to osteotomy plane<br>▪ Number of screws inside the bone<br>▪ Bi-cortical solutions preferred |

### 2.3.2. Fitness functions and constraints

Genetic algorithms employ minimization/maximization functions for the quantitative evaluation of the optimization parameters $\vec{x}$ during the optimization process. These functions are known as fitness functions and their definition is always challenging as they must be capable of providing a quantitative measurement of the quality of the different optimization objectives. The design of these fitness functions is a very challenging task, as we must consider not only mathematically solid functions that can be used by an optimization algorithm, but we must also guarantee that these functions reflect the clinical needs. For example, we cannot measure the quality of the osteotomy cut by simply measuring the inclination angle of the cut, because the bone contact and the bone protrusion of the generated fragments are of more clinical relevance than the orientation of the plane. In this section, we explain the design of the 4 fitness functions that we have developed based on clinically relevant measurements.

**Control of bone reduction alignment**

From the clinical point of view, some bone regions (i.e., joint regions) are more important to be precisely aligned to the reconstruction target than others. Moreover, in some cases, some bone regions inevitably deviate due to the pathology. We have developed a landmark-based error measurement to fine-control the reduction alignment by landmark regions $LP_l$ and $LR_l$ of the pathological bone and the reconstruction, respectively (**Figure 3B**). A weight $w_l$ between 0 and 1 was assigned to all points of $LP_l$, with $\sum w_l = 1$. Landmark regions located on the joint surface (i-v; **Figure 3B**) were given a bigger weight due to the importance of the joint for the reconstruction accuracy, which yields the following weight distribution $\{w_1, \ldots, w_5 = 0.18, w_6, w_7 = 0.05\}$. In this way, the quality of the reduction alignment is controlled by

$$f_1 = \frac{1}{|\{LP\}|} \sum_{l=1}^{|\{LP\}|} w_l \sqrt{\frac{1}{N} \sum_{i=1}^{N=50} (A_f\, p_i - q_i)^2} \text{ where } p_i \in LP_l, q_i \in LR_l \quad \textbf{(Eq. 6)}$$

which is the average weighted $RMSE$ among the Euclidean distances between all $A_f$-transformed points $p_i$ of the pathological landmark regions and the points $q_i$ of the landmark regions of the reconstruction target. $|\{LP\}|$ is the cardinality of the set $LP$. The points are in correspondence, meaning that $p_i$ and $q_i$ have the same position relative to their landmark centers. $A_f$ is constructed from $R_f = (\varphi_x, \varphi_y, \varphi_z)$ and $\vec{t}_f = (t_{fx}, t_{fy}, t_{fz})$ obtained from chromosome $\vec{x}$. To guarantee a clinically acceptable reduction alignment, $f_1$ must be within a user-defined error boundary ($\leq 0.5\, mm$). With this approach, it is possible to control the precision for specific regions on the bone. It allows not only a precise control of the reduction, but also the introduction of small deviations to the reconstruction target. Such deviations can support finding a better overall solution in case of contradicting targets by accepting small errors in the reduction alignment in favor of improving other objectives. For example, in case that the position of the fixation plate with respect to the $A_f$-transformed bone fragment might not be within the acceptable solution range, the algorithm could allow a larger transformation error, e.g., 0.5 mm, and this small variation could already generate solutions with better fitting positions of the fixation plate.

### **Control of osteotomy plane**

The parts of the bones that deviate from the anatomical target and that entail a surface gap among the fragments, are often referred to as bone protrusion (**Figure 7A**). The optimization of bone protrusion is very challenging because it is dependent on both, the osteotomy plane and the reduction alignment. We have decided to control the osteotomy cut position and direction implicitly by minimizing the bone protrusion among the generated fragments, in order to have a measurement that can be easily described to surgeons. As a positive side effect, the explicit optimization of the osteotomy plane is avoided, which has been proven to be challenging (Athwal et al., 2003; Schkommodau et al., 2005). The bone protrusion function $BP$ is defined as

$$BP(F, R) = \frac{1}{|F|} \sqrt{\sum_{p_i \in F} W(lp, p_i)^2 - \delta(W(lp, p_i), W(lp, R))^2}, \quad \textbf{(Eq. 7)}$$

where $F$ is the point set of a bone fragment and $R$ is the reconstruction target. Windows thickness $\vec{t}$ is set to $t_y = 4\, mm$ and $t_x$ and $t_z$ large enough to enclose $F$ and $R$ along $\vec{CS}_x$ and $\vec{CS}_z$.

In the optimization process, fitness function $f_2$ is defined as

$$f_2 = \frac{BP(P^{prox}, R) + BP(A_f P^{dist}, R)}{2}. \quad \textbf{(Eq. 8)}$$

An illustration of the bone protrusion calculation for $P^{prox}$ is shown in **Figure 7B-C.**

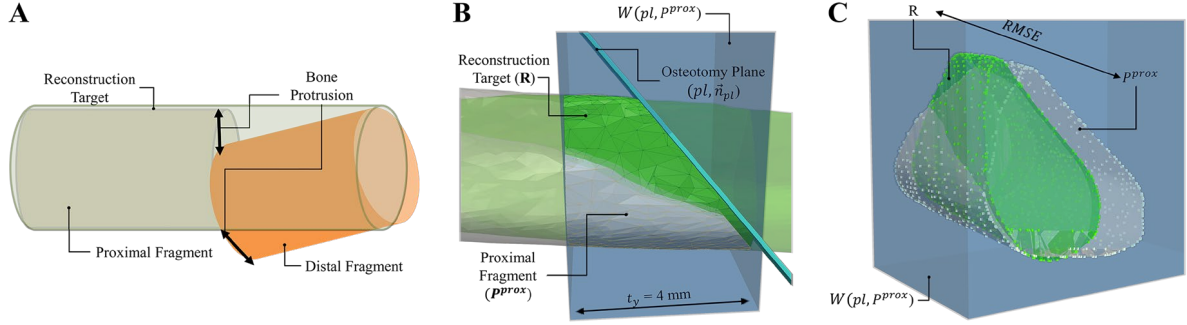

**Figure 7: Example of bone protrusion calculation.** (A) Simplified drawing of the geometrical representation of bone protrusion after osteotomy cut and realignment. Gray cylinder represents the proximal fragment, orange cylinder represents the reduced distal fragment. (B) Graphical explanation of the window function for fragment $\mathbf{P^{prox}}$; window function $W(lp, P^{prox})$ is shown in blue, osteotomy plane is shown in cyan, the proximal fragment is shown in white and the reconstruction target is shown in transparent green. (C) Calculation of the bone protrusion $BP(P^{prox}, R)$, which can be interpreted as the calculation of the RMSE between the point sets $\mathbf{P^{prox}}$ (white points) and $R$ (green points) subject to the function window $W$ (see **Eq. 7**).

## Control of plate position

Stability of the bone reduction and, consequently, successful healing, depends on an appropriate implantation of the fixation plate. In general, a minimal distance between fixation plate **FP** and bone surface is desirable. However, in orthopedic surgeries, the position of the fixation plate is often constrained by the approach site of the surgery (Klausmeyer and Mudgal, 2014) and by the intrinsic anatomy of the bones (muscle and ligament attachments and joints). To integrate this information into our algorithm, we have defined clinically feasible plate location regions according to all standard approach sites (Klausmeyer and Mudgal, 2014). These regions were annotated by clinical experts on an average mean model, generated by a statistical shape model (SSM) of the forearm (Mauler et al., 2017; Sepasian et al., 2014) (**Figure 8A**). The annotation was performed using the Scalismo mesh painting tool (Graphics and Vision Research Group, University of Basel, Switzerland). Region transfer from the mean model to each patient-specific model was achieved through a model fitting registration algorithm described by (Lüthi et al., 2018). Let $BR^i \in P$ the i-th feasible bone region fitted to the pathological bone that can be retrieved by a function $BR(P, i)$.

As a first pre-processing step, the fixation plate is coarsely aligned to each $BR^i$ of $P$ through ICP methods (Besl and McKay, 1992) and used as input for the optimization algorithm. In a second preprocessing step, an automatic identification is performed to determine plate model points $FP^*$ that should be brought into contact with the bone surface. To this end, an inherent coordinate system $\vec{CS}^{FP}$ for the fixation plate is calculated using a principal component analysis (PCA; (Jolliffe, 2011)) of FP. The eigenvectors of the PCA correspond to the Euclidian coordinates of the $\vec{CS}^{FP}$ and the origin $cs_o^{FP}$ is given by the mean of $FP$. Without loss of generality, let the positive direction of the $\vec{CS}_z^{FP}$ axis point towards the undersurface of the plate. A point of $p_i \in FP$ is included in point set $FP^*$ if $(p_i - cs_o^{FP}) \cdot \vec{CS}_z^{FP} > \varepsilon \, t_z^{FP}$, where $t_z^{FP}$ is the thickness of the fixation plate along $\vec{CS}_z^{FP}$. We empirically set $\varepsilon = 0.85$, which worked for all plate types. An example of the identified $FP^*$ point set is shown in **Figure 8B**.

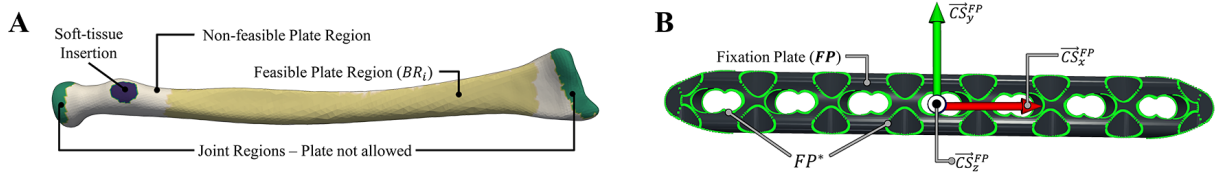

**Figure 8: Feasible plate regions.** (A) Feasible plate regions $BR_i$ (light yellow), encoded in the mean model of a radius SSM. The remaining colored regions are shown as indication of non-feasible plate regions (B) Surface points $FP^*$ (in green) of the edge of the plate that should be in contact with the bone after registration. Coordinate axis $\overrightarrow{CS_z^{FP}}$ is shown in black and is oriented towards the undersurface of the plate.

On each optimization step, the average distance between the fixation plate and the proximal fragment $P^{prox}$ is calculated as

$$D^{prox} = \frac{1}{|FP^*|} \sum_{\substack{i=1,\dots,|\{BR_i\}| \\ j=1,\dots,|FP^*|}} \left\| \delta\left(A_p\, p_j, BR(P^{prox}, i)\right) - p_j \right\|, \quad \text{where } p_j \in FP^* \text{ (Eq. 9)}$$

where $|FP^*|$ is the cardinality of $FP^*$, $A_p$ is the 4x4 homogenous transformation matrix of the plate obtained from chromosome $\vec{x}$, $BR(P^{prox}, i)$ are the points of feasible region $i$ contained in $P^{prox}$, and $\delta$ is the already defined nearest neighbor function of **Eq. 3**. Similarly, the distance between the fixation plate $FP$ and the distal fragment $P^{dist}$ is obtained by

$$D^{dist} = \frac{1}{|FP^*|} \sum_{\substack{i=1,\dots,|\{BR_i\}| \\ j=1,\dots,|FP^*|}} \left\| \delta\left(A_p\, p_j, A_f BR(P^{prox}, i)\right) - p_j \right\|, \quad \text{where } p_j \in FP^* \text{ (Eq. 10)}$$

where $A_f$ is the 4x4 homogenous transformation matrix of the distal fragment obtained from chromosome $\vec{x}$. The average of both distance values is used for controlling the position and orientation of the plate, encoded into fitness function $f_3$ as

$$f_3 = (D^{prox} + D^{dist})/2 \text{ (Eq. 11)}$$

### **Control of screw position and orientation**

The bone density of the patient bone considerably influences screw stability, as cortical and trabecular bone distribution differs along the bone (Kubiak et al., 2006; Reitman et al., 2004). The lengths of the screws play another important factor, as too short or too long screws might cause complications or injuries (Blecha et al., 2005; Spahn, 2004; Wall et al., 2012). Also, the position and direction of screws with respect to the bone density considerably affect the stability of the fixation (Dougherty et al., 2008; Weninger et al., 2010). Therefore, the bone density distribution will be considered for defining the screw positions.

We have developed a method for generating a patient-specific bone density mask based on the original CT image $I^{Orig}$ containing the patient-specific normalized Hounsfield values ($I^{Orig}$) (**Figure 9A-B**). Let $\Phi_I(p_i)$ be a transformation function transforming the 3D coordinates of a point $p_i$ to image space coordinates $(I_x, I_y, I_z)$ with respect to an image $I$. The outer cortical surface of the pathological bone model $P$ (orange overlay in **Figure 9B**) is rasterized into an empty image $I^{STL}$ of same size as $I^{Orig}$ using the method described by (Pineda, 1988). A grey value of $1000\, max_{(x,y,z)}(I^{Orig}(x, y, z))$ is assigned to each rasterized voxel. Next, a dilation of size 1 pixel is applied to $I^{STL}$ using a standard dilation kernel and the weighted bone-density mask shown in **Figure 9C** is defined

$$M^W(x, y, z) = \max\left(M^{Orig}(x, y, z), I^{STL}(x, y, z)\right), \quad \text{(Eq. 12)}$$

which returns the values of $I^{STL}$ for $(x, y, z)$ lying on the cortical layer, values of $M^{Orig}(x, y, z)$ lying inside the pathological bone (trabecular bone) and 0 otherwise. The masking process was performed in Matlab (MATLAB, 2017) using the Iso2Mesh toolbox (Fang and Boas, 2009).

In the objective evaluation of the optimization process, penetration points $in_{s_i}$ and $out_{s_i}$ (**Figure 9D-E**, cyan points) are calculated between each screw $s_i$ and the bone fragments in reduced position. The penetration points are found by casting a ray formed by screw center $A_p \vec{t}_{s_i}$ (**Eq. 4**, with $R_{s_i} = I_{4x4}$) and screw axis $\vec{h}_{s_i}$ and calculating the intersection with the bone as described in (Mount and Arya, 1998; Schneider and Eberly, 2002). Afterwards, a sampling between $in_{s_i}$ and $out_{s_i}$ is performed in 3D space along the screw axis using line equation $L_{s_i}(\lambda_{s_i}) = in_{s_i} + \lambda_{s_i}(out_{s_i} - in_{s_i}) \mid \lambda_{s_i} = \{0, 0.1, \ldots, 1\}$ (**Figure 9D**, cyan and red spheres). The sampling approach in 3D space is computationally more efficient than evaluating the entire density in the image space. The average weighted density value for each screw $s_i$ is obtained by

$$DV_{s_i} = \frac{1}{|\lambda_{s_i}|} \sum_{\lambda_{s_i} = \{0, 0.1, \ldots, 1\}} M^W\left(\Phi_{M^W}\left(l_{s_i}(\lambda_{s_i})\right)\right). \quad \text{(Eq. 13)}$$

The position and direction of the screws is controlled by the average of all the weighted density values

$$f_4 = \frac{1}{N_{s_i}} \sum_{i=1}^{N_{s_i}} DV_{s_i} \quad \text{(Eq. 14)}$$

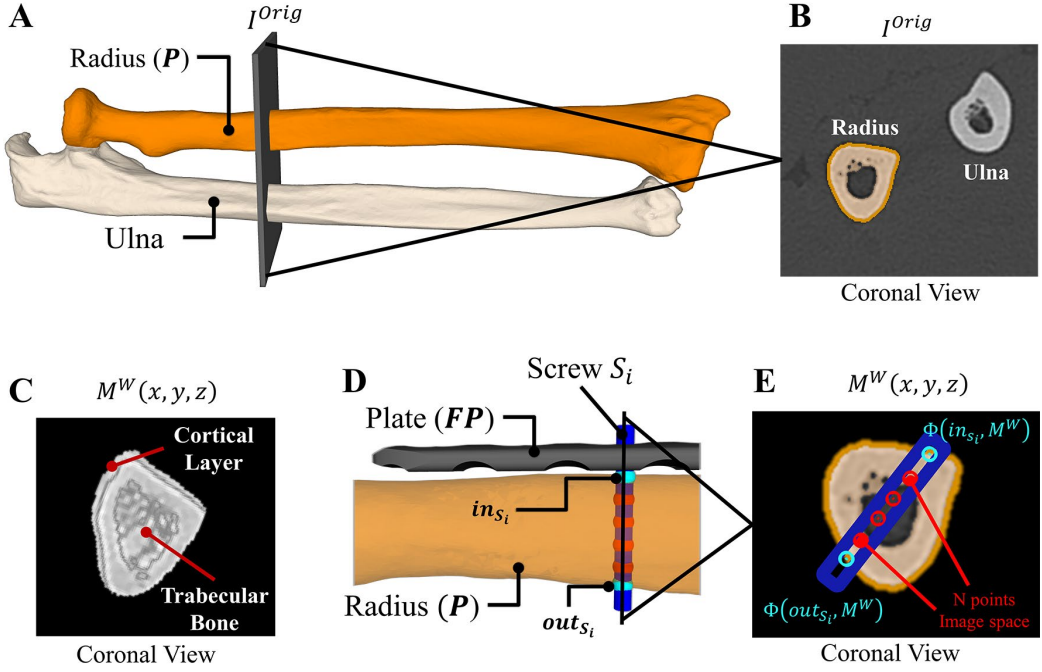

**Figure 9: Fitness calculation for screw position and direction using patient-specific bone density information.** (A) 3D models of the pathological radius (in orange) and ulna (in white); grey plane indicates a CT slice. (B) Slice of $I^{Orig}$ in coronal view. The radius is indicated by the semi-transparent orange mask and the outer cortical layer by the orange line. (C) Resulting weighted bone-density mask $M^W(x,y,z)$. (D) Sampling calculation of insertion points for Screw $S_i$; the cyan spheres represent $in_{S_i}$ and $out_{S_i}$ and the red spheres represent the sampled points along the screw axis. (E) Coronal view of $M^W(x,y,z)$ showing the corresponding points of $L(\lambda_{S_i})$ in image space.

### 2.3.3. Core Algorithm

The core optimization algorithm is a modified version of the multi-stage multi-objective genetic optimization approach previously described in our work (Carrillo et al., 2017), capable of handling multiple contradicting objectives and nonlinear constraint. Once each planning objective has been parameterized using the real-valued chromosome $\vec{x}$, an optimal Pareto set $Y$ of planning solutions can be found by

$$Y = \{y \in \mathbb{R}^m : y = \min_{\vec{x} \in X} (f_1(\vec{x}), f_2(\vec{x}), \dots, f_m(\vec{x}))\} \quad | \quad f_i: \mathbb{R}^n \to \mathbb{R}^m, \quad \textbf{(Eq. 15)}$$

where $\mathbb{R}^m$ is the solution space generated by the fitness function $f_i(\vec{x})$, $\mathbb{R}^n$ being the parameter space and X the set of optimal chromosomes generated by the algorithm. In our case, $\mathbb{R}^n = \mathbb{R}^{|\vec{x}|} \geq \mathbb{R}^{18}$ depending on the number of screws for each plate, and $\mathbb{R}^m = \mathbb{R}^4$ where $m$ is the number of fitness functions.

However, standard NSGA-II (Deb et al., 2002) is only able to find symmetrically optimized solutions with equally weighted optimization objectives on the solution space $\mathbb{R}^m$. In our optimization problem, each objective must have a different importance, according to its clinical relevance. Based on (Carrillo et al., 2017), we have applied the following weighting schema:

$$w(\vec{x}) = \sum_i e^{-k_i \vec{x}/e^{k_i}}. \quad \textbf{(Eq. 16)}$$

The weighting function $w$ increases the sparsity of solutions $Y$ around the utopia point (Miettinen, 1999). The sparsity of $f_i(\vec{x})$ along the Pareto set is controlled by the constant $k_i$, which represent the complementary percentage of the desired sparsity. For instance, if $k_i = 30$, $100 - k_i = 70\%$ of solutions are to be found closer to the utopia point. We have defined, together with surgeons, the following optimal weighting schema: Reduction alignment as the objective with the highest priority, followed by the osteotomy plane, the plate position and the screw purchase, respectively.

An initialization and automatic verification of constraints and boundaries is done previous to each optimization stage. The distribution of the optimization process into different stages allows handling contradictive objectives by giving different importance on each stage to different objectives.

**Initialization Stage 1.** An initial population $X^1$, where the superscript denotes the optimization stage, of 200 chromosomes is randomly generated among the parameter range of chromosome $\vec{x}$ (**Eq. 5**). At each stage, we optimize over all parameters (all objectives) but considering only a subset of fitness functions. A summary of the algorithm parameters is given in **Table 2**.

**Stage 1.** In this stage, the reduction alignment ($f_1$), the osteotomy plane ($f_2$), and the position of fixation plate ($f_3$) are optimized. Our algorithm is run with 3 objectives and weighted towards $f_1$ ($k_1 = 20$, **Eq. 16**) and $f_2$ ($k_2 = 40$, **Eq. 16**). This strategy permits to obtain solutions for the reduction alignment and bone protrusion close to the utopia point, but allowing a larger freedom on the plate position ($k_3$ is set to 100, i.e., sparsity is unaffected). By forcing the algorithm to remain in the solution space where the reduction alignment ($f_1$) and the osteotomy plane ($f_2$) are close to the utopia point, contradictive solution that might engender the quality of $f_1$ or $f_2$, are automatically neglected by the evolutionary process of the GA. The resulting Pareto set is stored in matrix $Y^1$ containing the best 70 solutions corresponding to the Pareto front. Accordingly, $X^{1*}$ represents the corresponding set of parameters yielding $Y^1$.

Table 2: Summary of algorithm parameters for each stage

| Parameter | Stage 1 | Stage 2 |
|---|---|---|
| Input Population | $X^1$ | $X^2$ |
| Max # Generations | 200 | 200 |
| Population Size | 200 | 200 |
| Number of objectives | 3 | 3 |
| Fitness Functions $f_i(\vec{x})$ | $f_1, f_2, f_3$ | $f_1, f_3, f_4$ |
| Weighting Schema ($K_i$) | $K_1 = 20, k_2 = 40$ | $K_3 = 40$ |
| Output | $Y^1, X^{1*}$ | $Y^2, X^{2*}$ |

**Initialization Stage 2.** The best 70 solutions from stage 1 are used for the initialization of the first generation of stage 2. To this end, the parameter range $X^2$ of the initial population of stage 2 are randomly initialized per component, such that each component value is constrained to lie between the maximum and minimum values obtained from $X^{1*}$. Thus, $\vec{x}_i^2 \in [min(X_i^{1*}), max(X_i^{1*})]$ where $i$ is the component i-th of the chromosome $\vec{x}$. The algorithm parameters of stage 2 are given in **Table 2.**

**Stage 2.** In this stage, we optimize the reduction alignment ($f_1$), the position of fixation plate ($f_3$) and the average screw purchase ($f_4$). Our algorithm is weighted towards $f_3$ ($k_3 = 40$, ***Eq.* 16**), to guarantee solutions with an optimal fixation plate alignment, and to avoid deteriorating the reduction alignment. The remaining fitness functions are not weighted ($k_1 = k_2 = 100$). The resulting Pareto set is stored in matrix $Y^2$, containing the best 70 optimal chromosomes corresponding to the Pareto front. Thanks to our weighing schema, $Y^2$ contains optimal solutions that are all within an acceptable clinical errors. $X^{2*}$ was ranked according to the solutions $Y^2$ by the best combined fitness among the 3 objectives of stage 2, yielding $XR^{2*}$. The final output $\vec{x}^{Sol}$ of the algorithm is obtained by taking the solution with the best plate alignment among the ranked list

$$\vec{x}^{Sol} = \min_{\vec{x} \in XR^{2*}} \frac{(f_3(\vec{x}) - min(f_3(\vec{x})))}{\left(max(f_3(\vec{x})) - min(f_3(\vec{x}))\right)}.$$

## 3. Results

### 3.1. Datasets

We have performed a qualitative validation (section **3.2**) and a quantitative evaluation (section **3.3**) on retrospective cases of malunited radii, which have been included in a large clinical trial about CA corrective osteotomy. From these consecutive cases, 36 cases were eligible according to the inclusion criteria given in **Table 3**. All 36 patients were treated at our orthopedics department between 2015 and 2017 and underwent navigated forearm osteotomy surgery through 3D preoperative planning and patient-specific instrumentation. Informed consent to use the patient data in an anonymized form for computer analyses was obtained, with ethical approvals KEK-ZH-Nr. 2015-0186 and BASEC-Nr. 2018-01608. The corresponding CT scans used for the generation of the 3D models were obtained according to a standard scanning protocol (slice thickness 1 mm; 120 kV; Philips Brilliance 64 CT, Philips Healthcare, The Netherlands). 3D preoperative planning of all patients were prospectively created in a manual fashion using a preoperative planning software (CASPA, Balgrist CARD AG, Switzerland) by the responsible hand surgeon, together with an expert planning engineer. These solutions were considered as the Gold Standard (GS). The baseline of the selected data is presented in **Table 3**.

**Table 3:** Inclusion and exclusion criteria for clinical and technical evaluation of the clinical cases

| Inclusion | Exclusion |
|---|---|
| ▪ Signed informed consent for data use<br>▪ Age ≥ 16 years old<br>▪ Radius osteotomy<br>▪ Availability of both, pathological and contralateral CT data<br>▪ Complete 3D preoperative planning including fixation plate | ▪ Incomplete CT data<br>▪ Revision surgery<br>▪ Intra-articular corrective osteotomy<br>▪ Multiple osteotomies on one bone |

**Table 4:** Baseline data of the 36 cases included in the study

| Age (years)* | _Mean_ | _SD_ | |
|---|---|---|---|
| | 33 | ±14 | |
| **Gender** | _Male_ | _Female_ | |
| | 11 | 25 | |
| **Affected Side** | _Right_ | _Left_ | |
| | 19 | 17 | |
| **Location Malunion** | _Distal Radius_ | _Radius Shaft_ | |
| | 14 | 22 | |
| **Osteotomy Type** | _Single Cut_ | _Closing Wedge_ | _Opening Wedge_ |
| | 11 | 8 | 17 |
| **Fixation Plate** | 9 different types | | |

\* years at time of surgery

## 3.2. Qualitative Validation

A preoperative planning solution (optimization solution; OA) was generated by our algorithm for each of the 36 consecutive cases in an automated fashion. The anatomical site and osteosynthesis implant were defined to be the same as in the GS solution for each case. A clinical survey was performed with 6 readers (2 senior hand surgeons, 1 board-qualified orthopedic surgeon, 1 resident surgeon and 2 expert planning engineers). In the survey, we asked each reader the question: "which of the two presented solutions would you implement in the surgery without further modification?". The two solutions (GS and OA) were presented to the readers in a blinded fashion and in random order. The assessment for each case is shown in **Figure 10A.** OA solutions were chosen as the better solution in 55% of the times (**Figure 10B**). Out of the 55%, surgeons' votes represented 38% and engineers' votes 17%. Similarly, for the 45% of the GS votes, 29% account for the votes of the surgeons and the rest 16% represent the votes of the engineers.

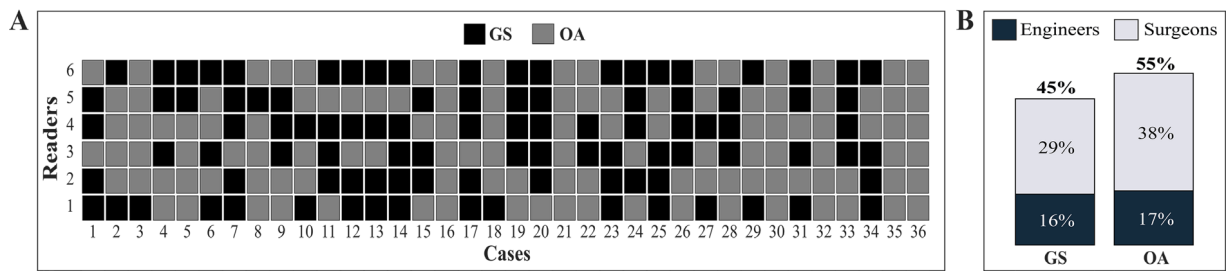

**Figure 10: Results of the qualitative validation**. (A) Voting of each reader and case. Each case was evaluated by 4 surgeons (readers 1-4) and 2 experienced planning engineers (readers 5-6). (B) Percentage distribution among the different voters for GS and OA solutions.

In **Figure 11** we present two examples where the OA solutions outperformed the GS solutions. In both cases, the reduction alignment and the position of the fixation plate obtained by the OA were clearly better than the GS solutions. Moreover, in both cases, the distance of the fixation plate with respect to the bone fragments has been decreased and undesired bone intersections have been solved by the OA. In these two cases, 100% of the readers favored the OA solutions.

There were also cases where the GS solutions were preferred by the majority of the readers. **Figure 12** shows two example cases where more than 50% of the readers favored the GS solutions. In the case shown in **Figure 12A**, surgeons preferred the GS solution due to a smaller bone protrusion with respect to the corresponding OA solution. In the case shown in **Figure 12B**, the GS solution has a better plate positioning, albeit a slightly worse reduction alignment compared to the OA solution.

In 8 out of the 36 cases, the GS solution corresponded to a closing wedge osteotomy (C1, C5-C7, C10, C18, C20, and C21) because planners failed to manually find a clinically acceptable single-cut osteotomy. For all of these 8 cases, the optimization framework was rerun and the algorithm converged towards a single-cut solution by applying the corresponding constraints. A single-cut osteotomy is the most preferred clinical implementation of a corrective osteotomy. **Figure 13** shows one of the OA single-cut solutions in comparison to the closing-wedge approach of the GS (**Figure 13A**). In the OA solution (**Figure 13B**), not only the osteotomy type was improved, but also the position of the fixation plate.

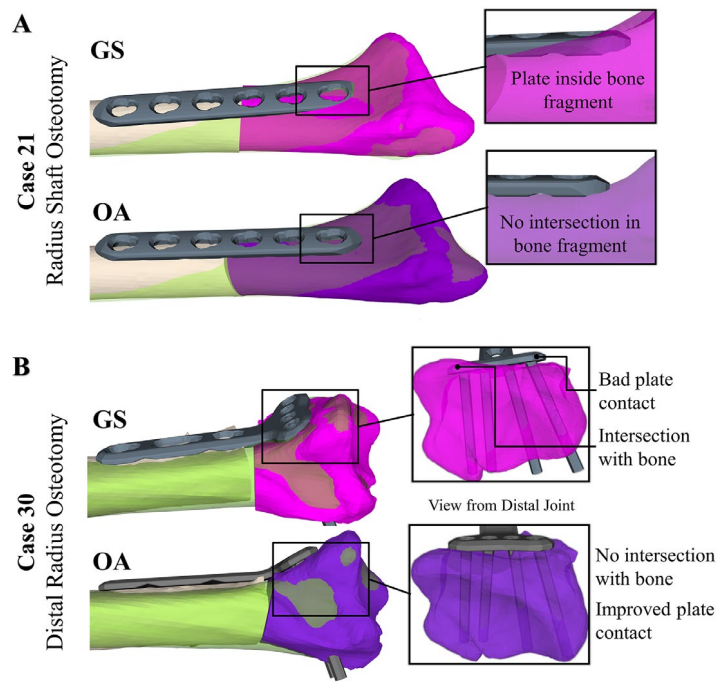

**Figure 11: Example cases for OA solutions that outperformed GS solutions.** The proximal fragment of the pathological bone is shown in white and aligned to the reconstruction target, shown in green. The reduced distal fragments of the OA and GS solutions are shown in magenta and purple, respectively. (A) Case 21 depicts a radius shaft osteotomy where the OA solution was able to solve undesired plate-bone intersections. (B) Case 30 shows a distal radius osteotomy. The OA solution was able to find a better positioning of the fixation plate avoiding bone intersections and improving the bone-plate contact.

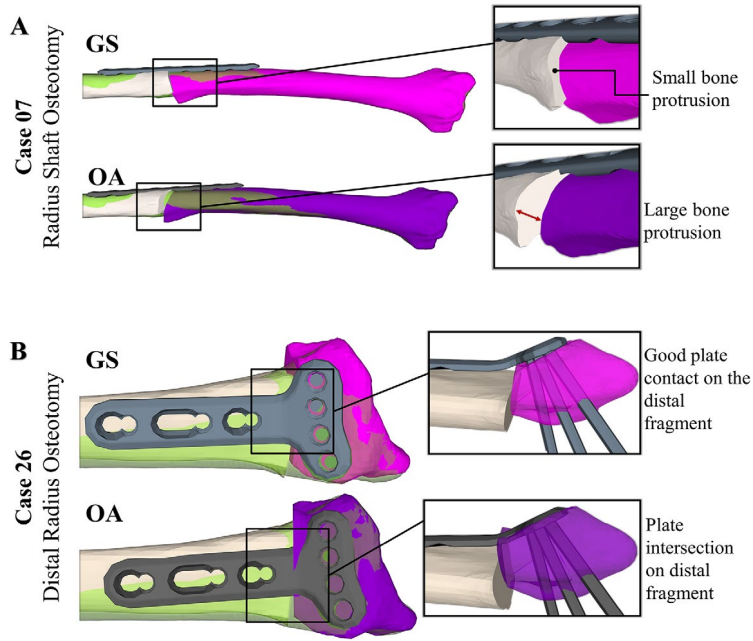

**Figure 12: Example cases where GS solutions were assessed to be better than OA solutions.** The proximal fragment of the pathological bone is shown in white and aligned to the reconstruction target, shown in green. The reduced distal fragments of the OA and GS solutions are shown in magenta and purple, respectively. (A) Case 07 depicts a radius shaft osteotomy where the bone protrusion was assessed to be better in the GS solution. The bone protrusion in the OA solution was larger in comparison to the GS, albeit a slightly better reduction alignment (B) Case 26 shows a distal radius osteotomy where the position of the plate was assessed to be better in the GS solution. In the OA solution, the plate-bone distance was optimized for the proximal fragment, resulting in a small bone intersection on the distal fragment.

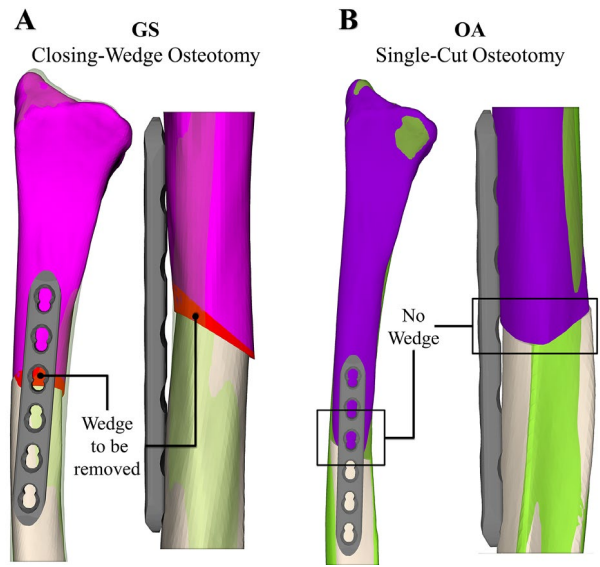

**Figure 13: Example of case 20, where the algorithm was able to find a single-cut osteotomy solution.** The proximal fragment of the pathological bone is shown in white and the reconstruction target is shown in green. (A) Closing-wedge solution generated by the state-of-the-art planning (GS); realigned distal fragment is shown in magenta and the wedge to be removed to complete the osteotomy is shown in red. (B) Single-cut solution generated by our approach (OA); realignment of the distal fragment is shown in purple. Position of the fixation plate was also improved with respect to the GS solution.

## 3.3. Quantitative Evaluation

Six different error measurements were evaluated for both, the GS and OA solutions, and for each of the 36 cases. Their average values, standard deviations, and ranges are reported in **Table 5**.

(a) Reduction error, measured by **Eq. 6**, between the transformed distal fragment and the reconstruction target. A reduction error <= 0.5 mm is desired clinically.

(b) Average distance between the fixation plate and the proximal bone fragment measured by **Eq. 9**.

(c) Average distance between the fixation plate and the distal bone fragment measured by **Eq. 10**.

(d) Average screw purchase of the proximal screws of the fixation plate, measured by **Eq. 14**. A larger screw purchase represents a better positioning of the screw.

(e) Average screw purchase of the distal screws of the fixation plate, measured by **Eq. 14**. A larger screw purchase represents a better positioning of the screw.

(f) Minimum distance between the screws of the fixation plate and the osteotomy plane measured by $min_{S_i} \; \|\vec{n}_{pl} \cdot (S_i - lp)\| / \|\vec{n}_{pl}\|$. A distance ≥2 mm is preferred.

**Table 5:** Mean, standard deviation and range for the 6 error measurements used in the technical evaluation of the optimization solutions compared to the gold standard.

| Measurement | Gold Standard (GS) | | | Optimization Algorithm (OA) | | |
| --- | --- | --- | --- | --- | --- | --- |
| | Mean | ± SD | Range | Mean | ± SD | Range |
| *Alignment Error (mm)* | 1.47 | 1.89 | 0.08 , 9.56 | 0.94 | 0.88 | 0.07 , 3.55 |
| *Distance Plate-Bone Proximal (mm)* | 0.83 | 0.67 | -1.04 , 2.64 | 0.57 | 0.44 | -0.80 , 1.59 |
| *Distance Plate-Bone Distal (mm)* | 0.88 | 0.66 | -1.05 , 2.03 | 0.64 | 0.44 | -0.88 , 1.62 |
| *Screw Purchase Proximal (unit)* | 1.4 | 1.02 | -0.86 , 2.90 | 1.98 | 0.74 | -0.22 , 3.34 |
| *Screw Purchase Distal (unit)* | 1.15 | 0.89 | -0.22 , 3.26 | 1.32 | 0.73 | -0.06 , 2.69 |
| *Minimum Distance Screw-Osteotomy Plane (mm)* | 4.91 | 3.05 | 0.45 , 9.59 | 5.75 | 2.53 | 0.80 , 12.34 |

We have also performed an evaluation with respect to each planning objective by assessing the error measures for both GS and OA solutions. **Figure 14** shows a set of box plots expressed as the relative improvement of the measures that were achieved by the OA solution. The blue baseline represents the GS value. In average, the alignment error was improved by 20% with respect to the GS among all the evaluated cases. The distance between the plate and the proximal and distal bone fragments has been improved by 26% and 31%, respectively. The average screw purchase for the proximal and distal screws was improved by 106% and 107%, respectively. Finally, an improvement of 93% in the average minimum distance between the fixation screws and the osteotomy plane was achieved by the OA solutions.

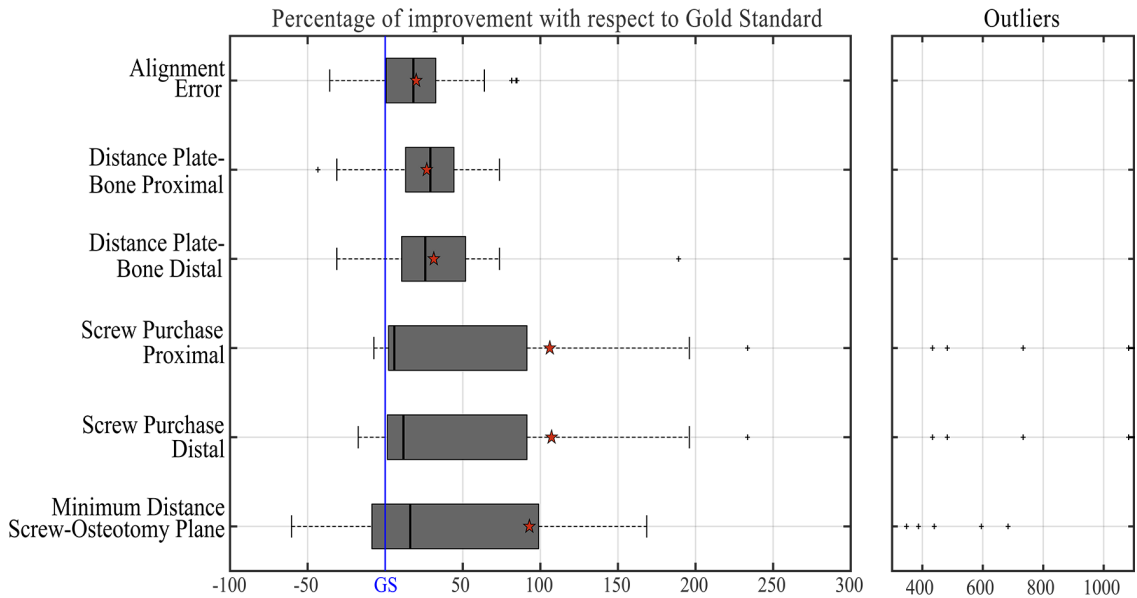

**Figure 14: Results of quantitative evaluation.** Box plot (3/2 interquartile range whiskers) of the relative improvement of OA solutions in relation to the GS (blue vertical bold line) for each of the 36 cases. The black bold line indicates the median and the red star indicates the average improvement for each error measurement. The black crosses indicate the outliers.

## 3.4. Runtime

The presented approach was implemented in Matlab R2017b (MATLAB, 2017). The average runtime of the algorithm comprising pre-processing, automatic diagnosis, verification and initialization of constraints, two optimization stages and generation of output solution was 85.38 min (±33.15 min).

## 4. Discussion

In the last decade, the use of computer-assisted 3D preoperative planning for orthopedic surgeries has increased significantly due to its higher precision and to the capability of treating more complex pathologies (Murase et al., 2008; Nagy et al., 2008). However, the great effort required for generating preoperative planning solutions with current state-of-the-art approaches poses a bottleneck in the treatment of corrective osteotomies. In this work, we have presented a computer framework for the fully automatic calculation of preoperative planning solutions of osteotomies. Our approach augments the skills of the surgeons by providing improved solutions with a reliable computational tool. To the best of our knowledge, this automated preoperative planning framework represents the first to have been developed to generate ready-to-use preoperative planning solutions for corrective osteotomies of the forearm

In orthopedic surgery, good surgical outcome is tightly associated with careful preoperative planning and precise surgical execution (Miyake et al., 2012b; Vlachopoulos et al., 2015; Vroemen et al., 2012). In order to demonstrate the clinical validity of our framework, we conducted a clinical study which confirmed the capabilities of the algorithm for generating ready-to-use preoperative solutions. The study showed that solutions generated by our algorithm were preferred in 55% of the time. In the remaining 45%, the quantitative evaluation confirmed still a high quality of the OA solution, with an error to the GS solutions of less than 1 mm for the reduction alignment, and less than 1.5 mm for the bone-plate distance.

One clear advantage of our framework is the ability of generating solutions in an automated fashion, deducting human calculation and planning times from the clinical setting, in contrast to state-of-the-art methods, which are associated with high effort and clinical costs (Fürnstahl et al., 2016). The calculation of 3D preoperative planning solutions in an automatic way would represent a time reduction of almost 35% of the human workload per case, decreasing therefore significantly the related costs. Our institution performs more than 250 corrective osteotomies each year using a commercial service, in which the 3D preoperative planning is performed in a manual fashion. If we consider average diagnosis and planning costs of USD 600 per case, our approach would save USD 150'000 of yearly clinical costs in our institution alone, while improving also the quality of planning and patient treatment.

Another advantage is the capability of calculating solutions with different trade-offs among the optimization objectives. Our algorithm can drastically improve one of the objectives by slightly decreasing the quality of other objectives. Such evaluations are not possible to obtain in a systematic way using the manual state-of-the-art approach (Bauer et al., 2017b; Vlachopoulos et al., 2015). Moreover, our method allows calculating single-cut osteotomies for arbitrary bone deformities including angular and translational deformities. In the presented case series, 8 single-cut OA solutions could be calculated where the real surgery was implemented with a closing wedge osteotomy. In 75% of these cases, readers validated the generated OA single-cut osteotomy solutions as a better preoperative planning than the original closing-wedge osteotomy proposed by a surgeon.

Concerning the surgical execution, the generation of complex solutions for preoperative planning comes also at the cost of an increase in the complexity of the intraoperative navigation. At our institution, we have developed a state-of-the-art technique for intraoperative guidance using PSI, which are 3D-printed individually for each case. These PSI allow a high level of precision for executing the preoperative plan in the surgery (Fürnstahl et al., 2016; Omori et al., 2014; Rosseels et al., 2018). An example of the generated PSI for intraoperative navigation of a distal radius osteotomy is shown in **Figure 15**. The accuracy of PSI navigation for corrective osteotomies has been previously evaluated, showing an improve in the postoperative outcome (Asada et al., 2014; Bauer et al., 2017a; Byrne et al., 2017; Caiti et al., 2018; Farshad et al., 2017; Gouin et al., 2014; Kunz et al., 2013; Michielsen et al., 2018; Omori et al., 2014; Roner et al., 2018; Rosseels et al., 2018; Schweizer et al., 2016; Vlachopoulos et al., 2015)

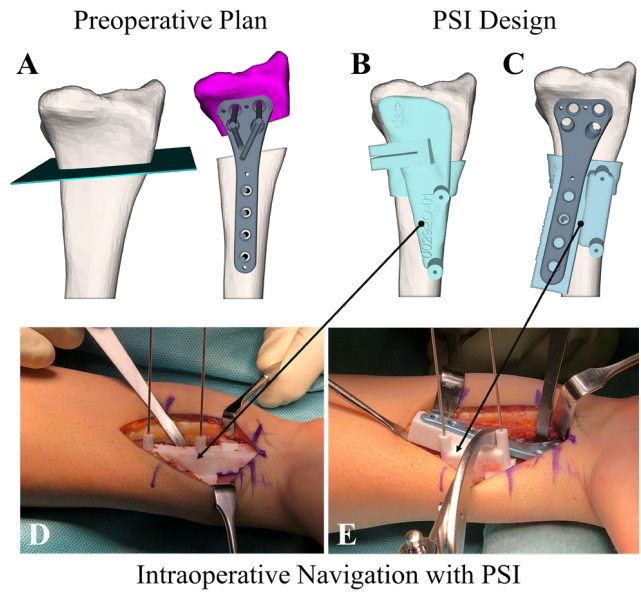

**Figure 15: Example of translation of preoperative plan to the intraoperative situation by means of 3D-printed PSI.** (A) Preoperative plan of a distal radius osteotomy. (B) PSI design for the osteotomy cut. (C) PSI design for the implant navigation and bone reduction. (D) Intraoperative navigation of the osteotomy cut. (E) Intraoperative navigation of the bone reduction and implant positioning. The black arrows indicate the PSI design that corresponds to each intraoperative situation.

The presented framework has some technical and clinical limitations. In the first place, the number of cases included in this study was not enough to provide statistical relevance to the quantitative and qualitative analysis performed. However, the observed trend shows a promising potential of the capabilities of the developed framework. The quality of the solutions generated by the optimization algorithm was very similar for every case to the ones of the Gold Standard. The decision towards one or the other was determined by very specific factors such as a better angle for the osteotomy cut or a better position for the fixation plate. Nevertheless, this is also the case when treating patients with the Gold Standard method as the surgeon has to decide which solution will be the one transferred to the intraoperative situation by means of PSI (Roner et al., 2018; Rosseels et al., 2018; Schweizer et al., 2013; Victor and Premanathan, 2013; Vlachopoulos et al., 2015; Wong et al., 2017)

In order to study the accuracy of surgical outcomes using automatic optimization methods, a prospective study with a statistical significant number of cases needs to be performed including a control group. Nevertheless, before such a clinical study can be approved and performed, the performance and safety of the algorithm needed to be validated. We have successfully achieved that in this paper through two different steps: (1) a quantitative evaluation showing that the algorithm is able to generate the expected solutions, optimizing all objectives, and (2) qualitative evaluation by a survey among surgeons and engineers showing that the algorithm produces solutions that the operating surgeon would execute in the practice.

Secondly, the optimization is not able to take into account clinical decisions made by the surgeons based on patient history, for example, an unusual reduction alignment due to soft tissue pathologies (Vlachopoulos et al., 2015), or a different fixation technique due to osteoporotic bones. Also, the presence of a deformity in the surrounding bones has a significant impact in the final reduction. In our algorithm, this influence was taken into account through the reconstruction target, however in future works, the influence of neighboring bones should be also taken into account as a part of the optimization framework. Furthermore, surgeons have intraoperative techniques to improve poorly fitting plates such as manual plate-bending or bone reshaping. Such actions cannot be foreseen by the algorithm. This might be also a reason why the GS solution was preferred in some cases even with a poor fitting of the fixation plate. Example of that can be seen for cases 7, 17, and 20, where the GS solution was preferred by 83% of the readers. In the future, patient-specific osteosynthesis and osteotomy implants such as the one described by Caiti et al. (2018) and Omori et al. (2014) could be integrated into the optimization framework.

Our approach is an important first step to turn automatic osteotomy planning into clinical practice. Due to the modularity of the algorithm, the approach could likely be extended to handle further anatomies (e.g., hip, humerus, clavicle, wrist bones) (Bugeau and Pérez, 2008; Chahla et al., 2016; Reising et al., 2013; Tschannen et al., 2016; Vlachopoulos et al., 2018). The positioning of the fixation plate and the calculation of the screw purchase could directly be adapted without significant efforts, but the calculation of the osteotomy plane and bone reduction would require the adaptation of the fitness functions and constraints.

Our algorithm does not include yet the possibility to support simultaneous osteotomies on multiple locations (Belei et al., 2007; Fürnstahl et al., 2016; Schkommodau et al., 2005), and it does not include either the capability to plan intra-articular osteotomies (Schweizer et al., 2013). However, we consider that the strategy could be adapted by including extra stages to the optimization pipeline, at the cost of increasing the calculation times due to the linear structure of the optimization stages. One possibility to avoid the increase of calculation times is to implement a parallel design for the optimization stages, also allowing the capability for a back-feedback mechanism of the entire optimization.

The presented approach is 18% faster in comparison to our previous work (Carrillo et al., 2017). However, the computational effort is still very high. One of the reasons for long calculation times is the implementation of the algorithm in a non-compiled programming language. Another possible source for the long calculation time is the lack of parallelization. We expect that the implementation of the entire pipeline in a compiled language and the use of graphical processing units (GPUs) for a customized parallelization of the approach will speed up the optimization framework to interactive rates.

## 5. Conclusions and Further work

The presented framework is able to generate clinically feasible preoperative planning solutions in an automatic fashion. The key idea of the approach is the use of a multi-staged, multi-objective optimization pipeline for the concurrent optimization of the planning objectives. We demonstrated that the clinical and technical quality of the solutions generated by the algorithm can be of the same quality as the solutions created by experienced surgeons. Moreover, in more than 50% of the cases, our algorithm was able to generate solutions that are hardly possible to be found within a reasonable time by the trial and error method of the Gold Standard. This shows the potential of the method to improve the quality of the preoperative planning by considering a larger range of feasible preoperative solutions. The algorithm does not intend to replace the expertise of surgeons in the process; our aim is to reduce unnecessary human work load. We seek to provide surgeons with automated tools for patient diagnosis and treatment, which can substantially decrease associated clinical times and costs, and generate better clinical outputs. Future works will focus on more flexible and less-invasive techniques for intraoperative guidance than patient-specific instrument (PSI) approaches such as augmented reality techniques in combination with learning-based methods for more comprehensive surgical guidance. Future solutions could also include a completely adaptive framework, were the optimal preoperative planning solution is adapted in real time according to the intraoperative situation.

**Conflict of interest**

None of the authors have any financial or personal relationships with other people or organizations that could bias their work.

**Acknowledgements**

This work has been funded through a Balgrist Foundation grant and a highly specialized medicine grant (HSM2) of the Canton of Zurich. We would also like to acknowledge the support and valuable input from the surgeons and planning engineers at the department of orthopaedics of Balgrist University Hospital.

**Vitae:**

**Fabio Carrillo** received his master degree in electronic engineering from Universidad Simon Bolivar in Caracas, Venezuela, with focus on automation. In 2015, he received his MSc degree in Robotics, Systems and Control from ETH Zürich, with a deep research interest in biomedical engineering. He joined the Computer Assisted Research and Development (CARD) Group at the University Hospital Balgrist in Zurich in September 2015, where he performs his doctoral studies, together with the Laboratory for Orthopaedic Biomechanics of the Department of Health Sciences and Technology at the ETH Zürich.

**Simon Roner** is an orthopaedic surgeon resident working as a research associate in computer-assisted surgery. He graduated from the medical school at Zurich University in 2012. After completion of Swiss board exams in orthopaedic surgery, he will pursue a specialization in hand and upper limb surgery.

**Marco von Atzigen** received his Master's degree in Mechanical Engineering from ETH Zurich with focus on Robotics, Systems and Control in July 2017. He joined CARD in October 2017 as a software developer working on machine learning and augmented reality. Since December 2018, he is pursuing his doctoral studies in deep learning for medical imaging and scene understanding between CARD, the spine surgery team at the University Hospital Balgrist Zürich and the Laboratory for Orthopaedic Biomechanics of the Department of Health Sciences and Technology at the ETH Zürich.

**Andreas Schweizer** graduated in Medicine from the University of Zurich. He received the Swiss Board Certification of Orthopaedic Surgery and Traumatology in 2003 and of Hand Surgery in 2005. He completed his MD thesis at Institute of Anatomy, University of Bern and his state doctorate at the University of Zurich. He is currently working as a consultant hand surgeon at the Balgrist University Hospital in Zurich, Switzerland and is specialized in 3D planning of orthopedic surgeries of upper extremity.

**Ladislav Nagy** graduated in Medicine from the University of Zurich in 1982. He received the Swiss Board Certification of Orthopaedic Surgery and Traumatology in 1990 and of Hand Surgery in 1992. He is titular professor of the faculty of medicine of the University of Zurich since 2011. He is the current chief of the hand surgery of the university hospital Balgrist and work as a consultant specialist in 3D planning of orthopedic surgeries of the hand.

**Lazaros Vlachopoulos** graduated in Medicine from the RWTH Aachen in Germany in 2004 and completed his MD thesis at the Institute of Human Genetics, RWTH Aachen. Afterwards, he completed his residency in orthopaedic surgery in Germany and Switzerland and received the Swiss Board Certification of Orthopaedic Surgery and Traumatology in 2012. In 2017 he received his PhD in medical image analysis from the ETH Zurich, Switzerland. He is currently working as a consultant orthopaedic surgeon at the Balgrist University Hospital in Zurich, Switzerland, specialized in computer-assisted orthopaedic surgery.

**Jess Snedeker** is Associate Professor of Orthopaedic Biomechanics, holding a professorial chair at both the ETH Zurich (Department of Health Sciences and Technology) and the Medical Faculty of the University of Zurich (Department of Orthopaedics). He heads the division of experimental research at the University Hospital Balgrist, and also serves as the chief scientific officer of the Balgrist Campus, designated in 2017 by the Swiss Secretariat for Education, Research, and Innovation as a "Research Infrastructure of National Relevance".

**Philipp Fürnstahl** received the MSc degree in technical mathematics and information procession from the Technical University of Graz, Austria, in 2005. In 2010, he received the PhD degree in medical image analysis from the ETH Zurich, Switzerland, for his research in the field of computer-assisted preoperative surgery planning. Philipp Fürnstahl is currently head of the Computer Assisted Research and Development (CARD) Group at the University Hospital Balgrist in Zurich, Switzerland, where he is involved in the development of patient-specific solutions for orthopaedics surgeries.